\RequirePackage{etoolbox}
\newbool{fullversion}
\booltrue{fullversion}
\documentclass[a4paper,USenglish,cleveref,thm-restate,pdfa]{lipics-v2021}

\pdfoutput=1 \hideLIPIcs  

\RequirePackage{etoolbox}

\providebool{fullversion}
\providebool{arxivversion}

\usepackage{preamble}
\usepackage{iris}
\usepackage{prob}
\usepackage{examples}
\usepackage{fontawesome5}

\ifbool{fullversion}{
\newcommand{\appref}[1]{\cref{#1}}
  \newcommand{\Appref}[1]{\Cref{#1}}
}{
\newcommand{\appref}[1]{the Appendix~\cite{fullversion}}
  \newcommand{\Appref}[1]{The Appendix~\cite{fullversion}}
}

\ifbool{arxivversion}{
\usepackage{caption}
}{
\usepackage[belowskip=-10pt,aboveskip=4pt]{caption}
\setlength{\intextsep}{15pt} }

\makeatletter
\ifbool{arxivversion}{}{
\let \MathparLineskip \mpr@lesslineskip }
\makeatother

\ifbool{arxivversion}{
  \newcommand{\vsquish}[1]{}
}{
  \newcommand{\vsquish}[1]{\vspace{-#1}}
}

\usepackage{tcolorbox}
\tcbuselibrary{skins,breakable}
  \newtcolorbox{result}{
    blanker,
    extras={interior engine=spartan},
    grow to left by=2pt,left*=0mm,
    grow to right by=2pt,right*=0mm,
    top=1mm,bottom=1mm,
    beforeafter skip balanced=0.1\baselineskip plus 2pt,
    breakable,
    colback=cyan!40
  }

\renewcommand{\thelang}{RandML}
\newcommand{\erisR}{Continuous-Eris{}}

\newcommand{\defer}[1]{\textcolor{lipicsLineGray}{#1}}

\newcommand{\mumax}{\mu_{\textrm{max}}}

\newcommand{\emax}{\langv{MaxU2}}

\newcommand{\eunif}{\langv{U}}
\newcommand{\pcts}{\mathcal{PC}}
\newcommand{\bdd}{\mathcal{B}}

\newcommand{\isR}{\langkw{IsReal}}
\newcommand{\infiniteTape}{\langkw{InfiniteTape}}
\newcommand{\stepsLeft}{\langkw{StepsLeft}}

\newcommand{\ForceNext}{\langv{ForceNext}}
\newcommand{\GetBits}{\langv{GetBits}}

\newcommand{\cmpU}{\langv{CmpU}}
\newcommand{\cmpUL}{\langv{C'}}
\newcommand{\RleHalf}{\langv{RleHalf}}

\newcommand{\NegExp}{\langv E}
\newcommand{\LaplaceSample}{\langv L}

\newcommand{\laplacepmf}[1]{\mathcal{L}_{#1}}
\newcommand{\toR}{\textrm{ToReal}}

\newcommand{\RofZ}{\langv{OfZ}}
\newcommand{\Radd}{\langv{Radd}}
\newcommand{\Rneg}{\langv{Rneg}}
\newcommand{\Rscal}{\langv{Rscal}}
\newcommand{\RofU}{\langv{RofU}}
\newcommand{\RofBZU}{\langv{RofBZU}}
\newcommand{\Rcmp}{\langv{Rcmp}}

\newcommand{\IsApprox}{\textrm{IsApprox}}
\newcommand{\ApproxTo}{\textrm{ApproxTo}}

\newcommand{\RealDecrTrial}{\langv R}
\newcommand{\HalfBernExpNeg}{\langv H}
\newcommand{\GeoTrial}{\langv G}
\newcommand{\IterTrial}{\langv I}
\newcommand{\GaussInt}{\langv Z}
\newcommand{\GaussReject}{\langv B}
\newcommand{\Gauss}{\langv N'}
\newcommand{\GaussSymm}{\langv N}
\newcommand{\Selector}{\langv S}
\newcommand{\SelectorZ}{\langv S'}
\newcommand{\FractionReject}{\langv A}
\newcommand{\IntChoice}{\langv C}

\newcommand{\distrof}[1]{\mu_{#1}}
\newcommand{\normof}[1]{\mathcal{N}_{#1}}

\newcommand{\Or}{\mathop{\,\|\,}}
\newcommand{\Mod}{\mathop{\%}}
\newcommand{\Unit}{()}

\newcommand{\additiveTR}{\text{\scalebox{0.7}{\faHourglass}}\hskip 0.1em}
\renewcommand{\TC}{\additiveTR}

\renewcommand{\epsilon}{\varepsilon}

\title{Verifying Exact Samplers for Continuous Distributions with a Discrete Program Logic}

\author{Markus de Medeiros}{New York University, USA}{mjd9606@nyu.edu}{https://orcid.org/0009-0005-3285-5032}{}
\author{Puming Liu}{NYU Shanghai, China}{pl2559@nyu.edu}{https://orcid.org/0009-0009-7095-564X}{}
\author{Kwing Hei Li}{Aarhus University, Denmark}{hei.li@cs.au.dk}{https://orcid.org/0000-0002-4124-5720}{}
\author{Alejandro Aguirre}{Aarhus University, Denmark}{alejandro@cs.au.dk}{https://orcid.org/0000-0001-6746-2734}{}
\author{Lars Birkedal}{Aarhus University, Denmark}{birkedal@cs.au.dk}{https://orcid.org/0000-0003-1320-0098}{}
\author{Joseph Tassarotti}{New York University, USA}{jt4767@nyu.edu}{https://orcid.org/0000-0001-5692-3347}{}

\authorrunning{M.\,de Medeiros, P.\,Liu, K.\,H.\,Li, A.\,Aguirre, L.\,Birkedal and J.\,Tassarotti}

\Copyright{Markus de Medeiros, Puming Liu, Kwing Hei Li, Alejandro Aguirre, Lars Birkedal, and Joseph Tassarotti} 



\relatedversion{A version including all appendices is availible online. } \relatedversiondetails[cite=fullversion]{Full Version}{https://arxiv.org/abs/2605.13526} 

\supplement{}
\supplementdetails[subcategory={}, cite={}, swhid={}]{Software}{https://zenodo.org/records/20114732}

\funding{
This work was supported in part by the National Science Foundation, grant no. 2338317, a Villum Investigator grant, no. 25804, Center for Basic Research in Program Verification (CPV), from the VILLUM Foundation, and the European Union (ERC, CHORDS, 101096090). Views and opinions expressed are however those of the author(s) only and do not necessarily reflect those of the European Union or the European Research Council. Neither the European Union nor the granting authority can be held responsible for them. Joseph Tassarotti is also employed by Amazon Web Services (AWS). This work does not reflect the views of AWS and is not connected to any products or services provided by AWS.
}

\nolinenumbers 

\EventEditors{Claudia Faggian and Joost-Pieter Katoen}
\EventNoEds{2}
\EventLongTitle{41st Annual Symposium on Logic in Computer Science (LICS 2026)}
\EventShortTitle{LICS 2026}
\EventAcronym{LICS}
\EventYear{2026}
\EventDate{July 20--23, 2026}
\EventLocation{Lisbon, Portugal}
\EventLogo{}
\SeriesVolume{380}
\ArticleNo{18}

\begin{document}

\maketitle

\begin{abstract}

  Most implementations of sampling algorithms for continuous distributions use floating-point numbers, which introduce round-off errors and approximations.
  These errors can be difficult to analyze, and can cause security issues when used in algorithms for differential privacy.
  An alternative is to use \emph{exact} sampling algorithms based on computable reals, which can lazily generate the digits of a continuous sample to arbitrary precision.
However, these algorithms are intricate, and implementing and using them involves a combination of semantically challenging language features, such as probabilistic choice, higher-order functions, and dynamically-allocated mutable state.

  In this paper we present \erisR{}, a higher-order separation logic for verifying the correctness of exact sampling algorithms for computable distributions.
To demonstrate \erisR{}, we verify the correctness of computable samplers for the uniform, Gaussian, and Laplace distributions, as well as a library for exact real arithmetic for working with generated samples.
  All of the results in this paper have been verified in the Rocq proof assistant.
\end{abstract}

\maketitle

\section{Introduction}
When computing with real-valued data, floating-point numbers are commonly used. Because of their bounded precision, operations on floating-point numbers inevitably introduce round-off errors.
For many applications, minor approximation errors are acceptable, and an expert who understands floating-point arithmetic can design algorithms to minimize such errors. 
However, for non-experts, the behaviors of floating-point arithmetic can sometimes be confusing, and in some applications, it is difficult to understand how much approximation error a computation introduces.
These floating-point errors can even lead to critical bugs or security issues.
For example, differentially private algorithms are often formally developed under the assumption that arithmetic is carried out exactly, and the presence of floating-point errors can lead to private information leakage~\cite{mironov2012significance}.

One alternative approach, which avoids round-off error, is to use algorithms for \emph{exact real arithmetic} over computable real numbers.
A number of different styles of exact real arithmetic exist, but the high level idea behind these approaches is to represent
a real number by a function $f$ which can be queried to get increasingly more precise approximations of the number.
Arithmetic operations on these numbers are higher-order programs, returning new functions which approximate the result of a real-valued operation in terms of approximations of its arguments.
This idea has its origins in constructive mathematics and the earliest days of computer science, and has been well-studied from both a theoretical perspective~\cite{DBLP:conf/lics/PottsEE97, DBLP:journals/tcs/Escardo96, DBLP:journals/iandc/Gianantonio96, DBLP:conf/mfcs/Simpson98, DBLP:journals/tc/Vuillemin90}  and an applied point of view, with several different implementations of exact reals as libraries or as built-in features of programming languages~\cite{menissiermorain:hal-02545650, DBLP:conf/pldi/LeeB90, DBLP:conf/pldi/Boehm20, DBLP:conf/cca/Muller00, DBLP:journals/entcs/BauerK08}.

What is less well-known is that by representing real numbers as functions, it is also possible to generate exact samples from continuous probability distributions.
Several algorithms exist for converting an infinite sequence of uniform Bernoulli samples into exact samples from the uniform distribution over the interval $[0, 1]$, the exponential distribution, and the Gaussian distribution, among others.
Using these algorithms, it becomes possible to do Monte Carlo simulations or other calculations with continuous random samples without approximations or round-off errors.

\begin{figure}
\begin{align*}
  \eunif{}\ \_ & \eqdef (\Alloc~\None,\ \defer{\textlang{tape}\ 1}) \\
  \ForceNext\ \ell\ \defer{\alpha}& \eqdef
  {\arraycolsep=1.4pt
    \begin{array}[t]{@{}rll}\multicolumn{3}{l}
    {\langkw{match}\spac \deref~\ell \spac\langkw{with}} \\| & \Some\ (b, \ell') &\Ra (b, \ell') \\
    | & \None & \Ra {\begin{array}[t]{@{}l}
        \Let b := \Rand\ \defer{\alpha}\ 1 in \\
        \Let \ell' := \Alloc~\None in \\
        \ell \gets \Some\ (b, \ell');; \\
        (b, \ell')
    \end{array}}\\\multicolumn{3}{l}{\langkw{end}}\end{array}} \\
  \GetBits\ \ell\ \defer{\alpha}\ N\ A & \eqdef
  \begin{array}[t]{@{}l}
    \If N = 0 then A \Else \\
    \Let (b, \ell') := \ForceNext\ \ell\ \defer{\alpha} in \\
    \GetBits\ \ell'\ \defer{\alpha}\ (N-1)\ (2*A+b)
  \end{array}
\end{align*}
\caption{A lazy exact sampler for the uniform distribution over the interval $[0, 1]$, and a client $\GetBits$.}\label{fig:u-impl}
\end{figure}

\cref{fig:u-impl} shows a simple implementation of an algorithm that lazily samples an exact value from the uniform distribution over the interval $[0, 1]$, written in an ML-like language (for now, ignore the code printed in a light \defer{gray} color).
The algorithm stores the sampled value as a mutable linked list of bits which represent the binary digits of the number that have been sampled so far.
The sampling function $\eunif{}$ simply creates an empty list, encoded as a reference cell containing $\None$, representing that no bits have yet been sampled.

Given a value $\ell$ sampled using $\eunif{}$, we can access approximations of the sampled value using $\GetBits{}$, which returns the first $N$ bits of the sample as a big-endian integer.
To calculate this approximation, $\GetBits{}$ traverses the linked list using a helper function $\ForceNext{}$, which takes a pointer $\ell$ to the next cell in the list as an argument.
The $\ForceNext{}$ function dereferences the pointer to check whether the next cell in the list exists, and if so, returns it.
Otherwise, if the next pointer contains $\None$, then we have reached the end of the list, so $\ForceNext{}$ samples a new Bernoulli value with the command $\Rand\ 1$, which returns $0$ or $1$ with equal probability.
It then appends the sampled value to the end of the linked list.
As we will see later~(\cref{sec:key-ideas}), using this $\GetBits{}$ function it is possible to convert this lazily sampled uniform deviate into a form that is compatible with existing libraries for computable real arithmetic, such as CReal~\cite{creal}.

Why is this algorithm correct, and in what sense do the programs $\eunif{}$ and $\GetBits$ sample a uniform value from $[0, 1]$?
At a high level, it is a standard fact from measure theory that an infinite sequence of Bernoulli samples, when interpreted as the binary digits of a real number, is equivalent to a uniform sample over $[0,1]$. 
But formally applying this argument to justify the code above is challenging.
First, the algorithm involves a combination of language features whose semantics are challenging to model and hard to reason about.
It directly combines mutable state, dynamically allocated pointers, and random choice.
If we wish to further reason about how these samples are used with constructive real libraries like CReal, then we must also consider a setting with higher-order functions.
Second, since the bits of the number are sampled lazily, at no point does our program actually obtain an infinite sequence of Bernoulli samples.
Yet we would still like to reason about $\eunif{}$ \emph{as if} it samples a uniform real value $r \in [0,1]$, in the sense that all calls to $\GetBits$ are just returning the digits of the binary expansion of $r$.

To the best of our knowledge, no techniques from prior work are able to verify an algorithm like the above.
While there has been extensive work on reasoning about probabilistic programs and verifying the correctness of sampling algorithms, prior work either restricts to discrete distributions~\cite{marionneau:samplers, sampcert, DBLP:journals/pacmpl/Bagnall0023}  or assumes the existence of primitives for sampling from continuous distributions~\cite{lilac, dash:lazy, basl, HirataMS22, SatoABGGH19} and works with non-computable real arithmetic.

To address this gap, we present \erisR{}, a higher-order separation logic which is capable of verifying the algorithm above, as well as other challenging exact samplers for continuous distributions.
This is possible due to a synchronicity between two concepts: continuous distributions are approximable by discrete distributions, and nonterminating programs are approximable by terminating traces.
In particular, by adding a variant of \emph{time receipts}~\cite{mevel19} to the Eris~\cite{eris} program logic, we can exploit this coincidence to obtain a program logic with continuous reasoning principles for computable, discrete sampling algorithms.

In summary, we present
\begin{itemize}
  \item A program logic for verifying computable continuous sampling algorithms using \emph{error credits} and \emph{time receipts},
\item A verified implementation of exact Gaussian, exponential, and Laplace sampling, and
  \item A formal proof that our Laplace sampler satisfies an accuracy bound used in the differential privacy literature.
\end{itemize}

The results in this paper have been mechanized~\cite{artifact} in the Rocq proof assistant, building on the Iris separation logic framework~\cite{irisjournal} and the Coquelicot real analysis library~\cite{coquelicot}.\footnote{Our development admits as axioms two standard facts about the Riemann integral that are missing from Coquelicot, but is otherwise fully verified. These axioms are Fubini's theorem for continuous functions and that the postcomposition of an integrable function by a continuous function is integrable.  }

 \section{Key Ideas}\label{sec:key-ideas}

\begin{figure}
  \begin{align*}
    \cmpUL\ \ell_1\ \defer{\alpha_1}\ \ell_2\ \defer{\alpha_2} & \eqdef {\begin{array}[t]{@{}l}
       \Let (b_1, \ell'_1) := \ForceNext\ \ell_1\ \defer{\alpha_1} in \\
       \Let (b_2, \ell'_2) := \ForceNext\ \ell_2\ \defer{\alpha_2} in \\
       \If b_1 < b_2 then -1 \Else \\
       \If b_1 > b_2 then 1 \Else \\
       \cmpUL\ \ell_1'\ \defer{\alpha_1}\ \ell_2'\ \defer{\alpha_2} \\
    \end{array}}\\
    \cmpU\ (\ell_1, \defer{\alpha_1})\ (\ell_2, \defer{\alpha_2}) & \eqdef \If \ell_1 = \ell_2 then 0 \Else \cmpUL\ \ell_1\ \defer{\alpha_1}\ \ell_2\ \defer{\alpha_2} \\
    \emax\ \_ & \eqdef {\begin{array}[t]{@{}l}
                      \Let x := \eunif{}\ \_ in \\
                      \Let y := \eunif{}\ \_ in \\
                      \If \cmpU\ x\ y < 0 then y \Else x
                      \end{array}}
  \end{align*}
  \caption{The maximum of two uniform $[0,1]$ samples. }
  \label{fig:MaxSampCode}
\end{figure}

To illustrate the key concepts of \erisR{}, we will verify the $\emax{}$ program in \cref{fig:MaxSampCode}.
The program $\emax{}$ draws two uniform deviates $x$ and $y$ using $\eunif{}$ and returns their maximum value.
To find the maximum, $\emax{}$ calls the helper function $\cmpU$, which determines the greater of two samples by comparing their bits in order, sampling to a higher precision if needed.
A standard exercise in probability theory is to prove that the value returned from the procedure is distributed over $[0,1]$ with probability density function $\mumax(x) \eqdef 2x$.
In this section we will demonstrate the key principles of \erisR{} by formally verifying this fact.

\subsection{Primer: Eris}\label{subsection:eris}

\erisR{} is an extension of Eris~\cite{eris}, a program logic for verifying the \emph{approximate correctness} of discrete probabilistic programs.
Eris is based on the Iris~\cite{irisjournal} separation logic framework, and uses techniques from Iris to support reasoning about programs which combine randomness, unbounded recursion, higher-order functions, and state.
The algorithms we verify in this paper use all of these features extensively, making Eris a good starting point.

The core construct in Eris is the \emph{error credit}, a separation logic resource $\upto{r}$ (for $r \in \real^{\ge 0}$) which can be ``spent'' to exclude an event whose probability is at most $r$.
The Eris adequacy theorem says that, for any predicate $P$ over values, a proof of $\hoare{\upto{\varepsilon}}{\langv{e}}{\Ret v. P(v)}$ implies that the probability of the program $\langv{e}$ terminating with a value that does not satisfy $P$ is at most $\varepsilon$.
Importantly, Eris is a partial correctness logic, so this statement implies nothing about the termination probability of $\langv{e}$.

Error credit assertions are used with the following rules:
\begin{enumerate}
  \item \emph{Spending}: $\upto{1} \vdash \bot$. The intuition is that a credit is an upper bound on the probability that a specification fails to hold, and an upper bound of $1$ on a probability is trivial.
  \item \emph{Splitting}: $\upto{\varepsilon_1 + \varepsilon_2} \dashv\vdash \upto{\varepsilon_1} \sep \upto{\varepsilon_2}$. This allows for splitting and joining error credits like other separation logic resources.
\alejandro{Conditioning is a very overloaded term, I would use something else to avoid confusion, e.g. distributing, weighted partitioning, etc}
\markus{Yeah, but I think we kind of want that to be the intuition, even if it's overloaded (allow a credit to depend on a random event). Perhaps you could read the new text in the remainder of this section and decide if it's still too overloaded?}
  \item \emph{Conditioning}:
    The value of an error credit can be conditioned on the outcome of a random sample drawn with the command $\Rand N$, provided that the expected value of the error credit is preserved:
\begin{equation}
      \quad \hoare{\upto{\mathbb{E}_{\unifd{N}}[\errfun]}}{\Rand N}{\Ret t. \upto{\errfun(t)} \sep t \in \{0, \ldots, N\}}
       \label{eqn:conditioning}
    \end{equation}
    The $\Rand N$ command returns an integer from the set $\{0, \ldots, N\}$ uniformly,
    and $\unifd{N}$ is the uniform distribution over that set.
\end{enumerate}
Error credits also inherit the rules of the underlying Iris separation logic, such as the frame rule.
Since Iris is an affine logic, excess error credits can be ``dropped'' or go unused.

To demonstrate how error credits work in practice, we will show how to spend $\upto{1/4}$ from a $\upto{1/2}$ error budget to exclude an event with probability $1/4$, as captured by the following specification:
\begin{equation}
  \hoare{\upto{1/2}}{\Rand 3}{\Ret t. \upto{1/4} \sep t \ne 0}.
  \label{eqn:erisexample}
\end{equation}
First, we use the \emph{splitting} rule to break our credit into two parts.
\begin{equation*}
  \upto{1/2} \vdash \upto{1/4} \sep \upto{1/4}.
\end{equation*}
The first credit will be framed across the call to $\Rand 3$.
For the second credit, set $\errfun(x) \eqdef \iverbr{x = 0}$, where the \emph{Iverson bracket} $\iverbr{P}$ has value 1 if $P$ holds, and 0 otherwise. Note that
$\mathbb{E}_{\unifd{3}}[\errfun] = 1/4$.
So, using the conditioning and frame rules we get
\begin{equation*}
  \hoare{\upto{1/4} \sep \upto{\mathbb{E}_{\unifd{3}}[\errfun]}}{\Rand 3}{t \ldotp \upto{1/4} \sep \upto{\errfun(t)}  \sep t \in \{0, \ldots, 3\}}
\end{equation*}
Now we perform case analysis on $t$.
In the cases where $t \ne 0$, we can continue the proof using the remaining $\upto{1/4}$ budget as desired.
When $t = 0$ we have that $\errfun(0) = 1$, so that $\upto{\errfun(0)} \vdash \bot$ by the spending rule,
allowing us to conclude.

\subsection{Correctness of Discrete Samplers via Error Credits}

As described above, Eris's adequacy theorem allows us to use error credits to upper bound the probability that a postcondition on the return value will fail to hold.
At first, it may appear that only being able to prove such upper bounds is a strong restriction on the kinds of probabilistic behaviors we can specify using Eris.
However, by considering more general classes of postconditions, one can completely characterize the distribution of values that a program may return.
In particular, \citeauth{marionneau:samplers} have shown that Eris satisfies a stronger soundness theorem that can be used to prove the correctness of sampling algorithms for discrete distributions.

The key insight is that the \emph{conditioning rule} for the $\Rand$ command captures the essence of the fact that $\Rand$ samples from the uniform distribution.
Turning this around, \authname{marionneau:samplers} show that to prove that an arbitrary program $\langv{e}$ samples from a discrete distribution $\mu$, it suffices to prove a specification about $\langv{e}$ which looks like a generalized version of~\eqref{eqn:conditioning}:
\begin{equation}
  \forall \rndvar \in \bdd(T), \hoare{\upto{\mathbb{E}_{\mu}[\rndvar]}}{\langv{e}}{\Ret t. \upto{\rndvar(t)}}
  \label{eqn:GenExpectation}
\end{equation}
where $\bdd(T)$ is the set of bounded functions $T \to \mathbb{R}^{\ge 0}$.
Informally, this specification states that $\langv{e}$ behaves like a random event with probability distribution $\mu$, in the sense that we can condition error credits on the outcome of $\langv{e}$ provided the average value stays the same.
To justify why a specification of the form in~\eqref{eqn:GenExpectation} suffices to prove that $\langv{e}$ generates samples with distribution $\mu$, we can consider two instantiations of the above specification for each value $t \in T$: first where $\rndvar$ is instantiated by $\overline{I}_t(x) \eqdef \iverbr{x \ne t}$ and second by $I_t(x) \eqdef \iverbr{x = t}$.
By applying the Eris adequacy theorem to the first instantiation, we get that the probability that $\mu(t)$ samples a value other than $t$ is at most $1 - \mu(t)$, and from the second instantiation we have that the probability that it samples $t$ is at most $\mu(t)$.
If $\langv{e}$ terminates with probability $1$, then by combining these inequalities, we have the probability that $\langv{e}$ returns $t$ is exactly $\mu(t)$, as desired.

When deriving specifications of the form in~\eqref{eqn:GenExpectation}, one is no longer reasoning about the probability of a particular error event.
Instead, the proof involves characterizing the expected value of an arbitrary bounded function $\rndvar$ applied to the return values of $\langv{e}$.
Error credits just become a way to systematically track how the steps of $\langv{e}$ transform that expected value.

A key benefit of this technique is that, not only does it show that a random sampler is correct, it does so in a way that can be used compositionally as part of a larger Eris proof.
In particular, once one has a specification of the form~\eqref{eqn:GenExpectation}, one can reason about $\langv{e}$ in the logic \emph{as if} it sampled from $\mu$.
\citeauth{marionneau:samplers} prove such specifications for a range of discrete sampling routines and show how they can be used for modular reasoning.

However, because their results and approach were restricted to discrete distributions, they cannot be used directly to verify samplers for continuous distributions like $\eunif$.
Our results show how to generalize their approach to handle continuous distributions.

\subsection{From Discrete to Continuous}\label{sec:keyideas:continuous}

To state the correctness of continuous samplers, we further generalize the pattern from~\eqref{eqn:GenExpectation}.
In place of expectations over discrete distributions, which are countable series, we instead use expectations over continuous distributions, as represented by a Riemann integral over a probability density function.\footnote{We use Riemann integrals instead of Lebesgue integrals because they suffice for the examples we consider here and are simpler to mechanize.}
For example, in the case of the uniform distribution over $[0, 1]$, the density function is just $p(x) = 1$, and so the specification pattern that we will prove for $\eunif{}$ has the form:
\begin{equation}
\begin{array}[t]{@{}l}
  \forall \rndvar \in \pcts([0, 1]),\, \hoare{\upto{\int_0^1 \rndvar(x)\,dx}}{\eunif\ \Unit}{\Ret v. \Exists r. \upto{\rndvar(r)} \sep \isR\,v\,r}.
\end{array}
  \label{eqn:advcomp}
\end{equation}
Here, $\pcts([0,1])$ is the set of bounded, piecewise-continuous functions $[0,1] \to \mathbb{R}^{\ge 0}$, and $\isR$ is a predicate stating that $v$ is a representation of the real number $r$ (we will define $\isR$ later in this section).
Piecewise continuity ensures that the Riemann integral exists.
Like in the discrete setting, equation~\eqref{eqn:advcomp} allows us to condition credits around a call to $\eunif{}$ \emph{as if} it returned a real number $r \in [0,1]$ uniformly at random.
Clients of $\eunif{}$ will be able to use this conditioning modularly as part of larger correctness and approximate correctness arguments (\cref{section:gauss} and~\cref{section:laplace}).
Finally, as we will see later in~\cref{section:soundness}, a specification of this form indeed captures what it means for the program to sample uniformly over the continuous distribution, in the sense that all approximants it returns are distributed with the correct probability.

But how can we prove such a specification in the first place?
There are several challenges.
First, as we previously discussed, at the time $\eunif$ returns, no values have actually been sampled yet, so how can we condition on some value $r$ being selected?
Second, the primitive conditioning rule for $\Rand$ is a discrete sum, yet the above involves an integral.
Finally, we need to formally define what it means for the function $v$ to represent the uniform deviate $r$ in Eris, with the $\isR$ predicate.

To address these issues, \erisR{} combines three ideas from previous logics.
First, \emph{pre-sampling tapes} from Clutch~\cite{clutch}, allow us to, as a proof device, pre-determine what the sampled values from future $\Rand$ commands will be.
With this we can pre-sample the bits that constitute the real number at the time that $\eunif$ is called.
However, Clutch's pre-sampling tapes are not enough on their own, as only allow for pre-sampling a \emph{finite} number of values in advance.
The binary expansion of the sampled value $r$ could have an infinite number of digits that are generated by calls to $\ForceNext$.

To overcome this, we exploit the fact that Eris is a partial correctness logic, and adapt the idea of \emph{time receipts} to internalize this partiality in the logic~\cite{mevel19}. Because of partiality, in any given execution that we reason about, we only need to consider up to some number $k$ of steps, where $k$ is arbitrary but bounded.
Since producing a new digit in the binary expansion takes at least one step, an execution of length $k$ cannot observe more than $k$ bits, so we only need to pre-sample $k$ values to know all of the bits that could eventually be seen in that execution.

Finally, by pre-sampling enough bits, we can get a Riemann sum with the conditioning rule that is arbitrarily close to the integral in~\eqref{eqn:advcomp}.
Using the thin-air error credit rule proposed by \citeauth{coneris}, we can logically pay for the discretization error that arises in passing from the sum to the integral.

The remainder of this subsection outlines these three ingredients and how they are used to derive the specification~\eqref{eqn:advcomp}.

\paragraph*{Pre-sampling tapes}
Pre-sampling tapes (or \emph{tapes}) were introduced in Clutch~\cite{clutch} as a proof device for representing knowledge about the random events in a program's future.
The operational semantics for tapes will be outlined in~\cref{subsection:language}, here we focus on the proof rules they enable.
In the logic, a tape is a separation logic resource $\alpha \hookrightarrow (N, \tape)$ representing the knowledge that all $\Rand$ commands marked with tape label $\alpha$ (denoted $\Rand\ \alpha\ N$) will draw their next $|\tape|$ samples from the finite list $\tape$, in order.
This is captured by the rule for $\Rand$, which says that the returned value will be the head of the tape.
\begin{equation*}
  \hoare{\alpha \hookrightarrow (N, n\cons\tape)}{\Rand\ \alpha\ N}{\Ret v. v = n \sep \alpha \hookrightarrow (N, \tape)}
\end{equation*}
We can presample a new value to the end of the tape, using the following rule, which allows us to condition error credits similarly to~\eqref{eqn:conditioning}:
\begin{equation}
  \infer
    {\hoare{\Exists n. \alpha \hookrightarrow (N, \tape \cdot n) \sep \upto{\errfun(n)} \sep P}
          {\langv{e}}
          {Q}}
    {\hoare{\alpha \hookrightarrow (N, \tape) \sep \upto{\mathbb{E}_{\unifd{N}}[\errfun]} \sep P}
          {\langv{e}}
          {Q}}
\label{eqn:presample1}
\end{equation}
By repeatedly applying this rule, we can pre-sample a finite number of values to a tape, that will then later be consumed by calls to $\Rand$.

The \defer{gray} code in~\cref{fig:u-impl} generates a fresh tape \defer{$\alpha$} for each uniform deviate returned by $\eunif{}$.
Intuitively we would like to presample an infinite sequence of bits representing some number $r$ onto this tape: doing so would mean that all future calls to $\GetBits$ will deterministically return approximations of a single real number $r$, and as a consequence, we could condition our error credits based on the value $r$ takes.
Note however that presampling tapes can only contain a \emph{finite} list of values.
We will now turn our attention to simulating infinite presampling operations using only finite tapes.

\paragraph*{Time Receipts}
Time receipts were introduced to separation logic by \citeauth{mevel19} to eliminate reasoning about program behaviors that would take an infeasibly long amount of time to occur (\eg overflowing a 64 bit counter).
In \erisR{} we adapt this idea to \emph{internalize} the fact that the logic is partial, so that we do not need to reason about diverging executions.
Specifically, we introduce an assertion of the form $\stepsLeft(k)$ establishing that $k$ is an upper bound on the steps left that we will reason about in the program, which must always be nonnegative, i.e.,
$k < 0 \to \stepsLeft(k) \vdash \bot$.
This is introduced using the following rule:
\begin{equation}
  \infer{\All k. \hoare{\stepsLeft(k) \sep P}{\langv{e}}{Q}}
        {\hoare{P}{\langv{e}}{Q}}
\label{eqn:intro-stepsleft}
\end{equation}
Read from bottom to top, applying this rule gives us $\stepsLeft(k)$ for some arbitrary $k$. Since we must prove the result for \emph{all} possible values of $k$, this ensures we consider all finite prefixes of possible executions.

Each step taken by the program generates a new \emph{time receipt} resource $\TC(1)$.
Time receipt resources combine additively like error credits, and can be spent to decrease the global runtime bound:
\begin{equation*}
  \stepsLeft(n) \sep \TC(m) \wand \stepsLeft(n-m).
\end{equation*}

With this in hand, we can construct a resource which provides the \emph{illusion} of an infinite presampling tape.
If we know we are reasoning about an execution that can be at most $k$ more steps, and the tape already has $k$ values on it, then any additional values can never be observed.
We define an assertion to represent a tape containing an infinite binary sequence, as represented by a function $f : \nat \to \{0, 1\}$. 
\begin{align*}
\infiniteTape \ \alpha \ f \ \eqdef \Exists k, \tape. \alpha \hookrightarrow (1; \tape) \sep \stepsLeft(k) \sep k < |\tape| \sep \All i. i < k \to \tape[i] = f(i)
\end{align*}

This assertion says that, under the hood, there is a real tape with at least $k$ samples which match the first $k$ values of $f$, and we have $\stepsLeft(k)$.
For this infinite tape assertion, we have a derived $\Rand$ rule
\begin{equation*}
  \hoare{\infiniteTape \ \alpha \ f}{\Rand\ \alpha\ 1}{\Ret v. v = f(0) \sep \infiniteTape\ \alpha\ (\Lam x. f(x+1))}
\end{equation*}
which says that executing $\Rand\ \alpha\ 1$ returns the first element $f(0)$ of the sequence, and returns a sequence that has been shifted by $1$.
The proof of this rule reasons by cases on whether the underlying tape assertion inside of $\infiniteTape$ has any samples left.
If it does, it pulls the first sample off of the tape, and also generates a time receipt letting us decrease the $\stepsLeft$ bound by 1.
When the tape is empty, the $\stepsLeft$ term will become negative, allowing us to conclude the proof by the time receipt principle instead.
Thus, externally, the specification makes it appear as if $\alpha$ contains an infinite number of bits matching the infinite bitstring $f$.

\paragraph*{Thin-Air Credits}
We complete our illusion by deriving a presampling rule that is capable of populating an $\infiniteTape$ with the binary expansion of a real number.
\begin{equation}
  \infer{
         \All r.
         \hoare{\upto{\errfun(r)} \sep \infiniteTape\ \alpha\ (\textrm{bin}\ r) \sep P}
               {\langv{e}}
               {Q}}
         {\hoare{\upto{\textstyle\int_0^1 \errfun(x) \,dx} \sep
            \alpha \hookrightarrow (1; []) \sep
            P
          }{\langv{e}}{Q}}
  \label{eqn:tapepresample}
\end{equation}
The function $\textrm{bin} : [0, 1] \to (\mathbb{N} \to \{0, 1\})$ gives a binary expansion of a real number, and it is not important which of the possible binary expansions this function picks.
Deriving this rule requires extending \erisR{} with one last ingredient: a rule for generating arbitrarily small error credits ``out of thin air'',
as proposed by \citeauth{coneris}:
\begin{equation}
  \infer{\hoare{\Exists \epsilon > 0. \upto{\epsilon} \sep P}{\langv{e}}{Q}}
        {\hoare{P}{\langv{e}}{Q}}
\label{eqn:thinair}
\end{equation}
This rule says that at any time, we can get access to some additional error credit of some arbitrary value. 
As \citeauth{coneris} show, this rule can be added without affecting the soundness of Eris through a limiting argument taking $\epsilon \to 0$.

To derive the specification~\eqref{eqn:tapepresample}, we first apply the rule~\eqref{eqn:intro-stepsleft} to get $\stepsLeft(k_1)$ for some $k_1$.
Next, we apply~\eqref{eqn:thinair} to get some credit $\epsilon > 0$, in addition to the $\int_0^1 \errfun(x)\,dx$ error credit we start with.
From Riemann integrability of $\errfun$, there exists $\delta > 0$, such that any Riemann sum over $\errfun$ with partition width smaller than $\delta$ will be within $\epsilon$ of $\int_{0}^1 \errfun(x)\,dx$.
Let $k_2$ be such that $1/2^{k_2} < \delta$. 
Set $k = \max(k_1, k_2)$. 
We now apply the basic presampling rule~\eqref{eqn:presample1} $k$ times to get $k$ bits on the tape $\alpha$.
If we interpret these bits as the first $k$ bits of the fractional representation of a real number, then unfolding the expected value sums from the repeated use of~\eqref{eqn:presample1}, we get a sum for the expected value of the resulting error credit that can be rewritten as
\[ \sum_{i = 0}^{2^k - 1} \errfun\left(\frac{i}{2^k}\right) \cdot \frac{1}{2^k} \]
This is a Riemann sum over $[0, 1]$ with partition width $1/2^k$, thus it is less than the $\epsilon +\upto{\textstyle\int_0^1 \errfun(x) \,dx}$ credit
that we have.
Since Eris is an affine logic that allows us to ``throw away'' excess error credits, we are therefore done.

With this infinite tape pre-sampling rule, we are finally able to derive~\eqref{eqn:advcomp}, the specification for $\eunif$.
We define the predicate $\isR\ v\ r$ to state that the heap location $v$ points-to a linked list of bits and an $\infiniteTape$, such that the bits on the list together with the bits on the infinite tape form a binary sequence for $r$.
Executing $\eunif{}$ gives us access to a heap location and empty tape, and we can apply~\eqref{eqn:tapepresample} to establish the $\isR$ predicate and the error credits in the postcondition of~\eqref{eqn:advcomp}.

\paragraph*{Summary}
In this subsection, we have seen how a few logical ingredients allow us to derive a continuous analogue of the specification style previously proposed by \citeauth{marionneau:samplers} for discrete samples.
In particular, extending Eris with time receipts allowed us to extend Clutch's finite pre-sampling tapes into a mechanism that behaves like an infinite tape.
Next we will explore how to use this specification in client code, using careful credit conditioning to verify algorithms based on uniform real samples.

\subsection{Verified Maximum Sampler}\label{subsection:emax}

Now we can now return to the verification of $\emax{}$ from~\cref{fig:MaxSampCode}.
We will do this in explicit detail as it illustrates the core techniques we will be making use of in our more complex examples.
Recall that the density of the maximum of two uniforms is given by $\distrof{\emax{}}(x) = 2x$.
We can state a continuous correctness theorem for $\emax$ as:
\begin{equation}
\begin{array}[t]{@{}l}
  \forall G \in \pcts([0, 1]),\,
  \hoare{\upto{\int_0^1 G(x) \cdot 2x\, dx}}{\emax\ \Unit}{\Ret v. \Exists r.
    \isR\ v\ r \sep \upto{G(r)}}.
\end{array}
  \label{eqn:max}
\end{equation}

In other words, given some $G$, and $\varepsilon_0 \eqdef \int_0^1 G(x) \cdot 2x\,dx$ error credits, we want to execute the body of $\emax{}$ and ensure that the return value of $\emax{}$ approximates a real number $r$, such that we end up with $\upto{G(r)}$ credits leftover.
To prove this, we will apply the specification for $\eunif{}$~\eqref{eqn:advcomp} at each of the uniform sampling statements, instantiating the function $\rndvar$ from that specification with appropriate credit conditioning functions.

For the first call to $\eunif$, we instantiate $\rndvar$ with the function $\varepsilon_1$ defined by
$\varepsilon_1(x) \eqdef \int_{0}^1 G(\max(x, y)) \,dy$.
Let $r_1$ be the real number we obtain in the post-condition of the rule.
Then, for the second call to $\eunif$ we instantiate $\rndvar$ with $\varepsilon_2$ defined by
$\varepsilon_2(y) \eqdef G(\max(r_1, y))$.
Call the result of this call $r_2$.

At this point there are two parts left to the proof.
The first is to show that the iterated integral we obtain from applying this rule twice with these choices of $\rndvar$ is equivalent to $\varepsilon_0$, the error credits we have in the precondition of~\eqref{eqn:max}.
We have
\begin{align*}
& \int_0^1 \int_0^1 G(\max(x, y)) \,dx \,dy \\
  &= \int_0^1 \int_0^1 \iverbr{x < y} G(y) \,dx \,dy + \int_0^1 \int_0^1 \iverbr{y < x} G(x) \,dx \,dy \\
  &= 2 \int_0^1 G(z) \int_0^1 \iverbr{w < z} \,dw \,dz \\
  &= \int_0^1 G(z) \cdot 2z\, dz \\
  &= \varepsilon_0
\end{align*}
The second part of the proof is to show that the code in fact computes the maximum $\max(r_1, r_2)$.
But this part of the reasoning involves no probability theory, as it amounts to reasoning about the bit-by-bit comparison done by traversing the linked-list of bits in the two numbers inside $\cmpU$ using $\ForceNext$.
Readers unfamiliar with Iris may be concerned that the recursive calls in $\cmpUL{}$ are not structurally decreasing in any argument.
Formally speaking, we verify $\cmpUL$ using the L\"ob induction rule from Eris.
\begin{mathpar}
\infrule[lab]{ht-rec}
  {\All \valB . \hoare{\prop}{(\Rec f x = e)\ \valB}{\propB} \vdash \hoare{\prop}{\subst{\subst{e}{x}{\val}}{f}{(\Rec f x = e)}}{\propB} }
  {\vdash \hoare{\prop}{(\Rec f x = e)\ \val}{\propB}}
\end{mathpar}
The L\"ob induction rule allows us to assume that the specification for $\cmpUL$ holds at every recursive call in its body.
In a call to $\cmpUL$, if the leading bits are different, then the real number starting with $1$ is greater than or equal to the one starting with $0$.
If the leading digits match, the program $\cmpUL$ will recurse and we conclude by the L\"ob induction hypothesis.

The proofs for subsequent more sophisticated sampling algorithms follow a similar pattern: first we choose the right instantiations of the function $\rndvar$ for each random sample the algorithm draws, and then we prove that the resulting iterated integrals are equivalent to the integral over the density of the distribution the algorithm is generating samples from.
Then, after the values are sampled, we are left with traditional reasoning about linked lists or other bit manipulations of the sampled values, for which separation logic is well-suited.
With \erisR{}, proofs of this format (including the more complex sampler in~\cref{section:gauss} and~\cref{section:laplace}) can be implemented formally inside the Rocq proof assistant.

\subsection{Interlude: Constructive Real Arithmetic}\label{sec:keyieas:creal}

While our lazy bit list representation of $[0,1]$ suffices for drawing and comparing uniform samples, clients of $\eunif{}$ such as the Gaussian or Laplace sampler will require us to represent samples over the entire real line.
Before we go deeper into these more complex sampling procedures, we will discuss a simple implementation of \emph{constructive real arithmetic}, which is compatible with the lazy bit lists of $\eunif{}$, and which we have verified in \erisR{}.

There are a number of options in the literature for representing lazily defined real numbers, such as signed bit sequences~\cite{Wiedmer1980}, continued fractions~\cite{DBLP:journals/tc/Vuillemin90}, or sequences of dyadic approximants~\cite{menissiermorain:hal-02545650}.
Our implementation is an adaptation of the CReal library~\cite{creal}, which is based on the latter technique.

We represent a real number $r$ using functions $s_r : \mathbb{Z} \to \mathbb{Z}$, which return integer approximations of $r$ to a specified accuracy.
Formally, we define a predicate $\ApproxTo(A, z, r)$ which says that $A \in \mathbb{Z}$ approximates $r$ to within a degree of precision specified by $z \in \mathbb{Z}$:
\begin{equation}
  \ApproxTo(A, z, r) \eqdef |A - r \cdot 2^z| \le 1.
  \label{eqn:crealbound}
\end{equation}
We say then that a function $s_r$ is a valid representation of $r$ if $\ApproxTo(s_r(z), z, r)$ holds for all $z$.
This condition ensures that as $z \to \infty$, the sequence $s_r(z)/2^z$ converges to $r$.

\begin{figure}
  \begin{align*}
    \RofZ\ z\ p & \eqdef z \ll p \\
    \Radd\ r_1\ r_2\ p & \eqdef
    \begin{array}[t]{@{}l}
      \Let z := r_1\ (p+2) + r_2\ (p+2) in \\
      (z / 4) + (z \Mod 4) / 2
    \end{array}\\
    \Rneg\ r\ p & \eqdef -(r\ p)\\
    \Rscal\ r\ z\ p & \eqdef r\ (p-z) \\
    \RofU\ u\ p & \eqdef \If p \le 0 then 0 \Else \GetBits\ u\ p\ 0 \\
    \Rcmp\ r_1\ r_2\ p & \eqdef
    \begin{array}[t]{@{}l}
      \Let n_1 := r_1\ p in \\
      \Let n_2 := r_2\ p in \\
      \If n_1 + 2 < n_2 then -1 \Else \\
      \If n_2 + 2 < n_1 then 1 \Else \\
      \Rcmp\ r_1\ r_2\ (p+1)
    \end{array}
  \end{align*}
  \caption{The \erisR{} constructive real library}
  \label{fig:creal}
\end{figure}

\Cref{fig:creal} depicts our implementation of a subset of operations from CReal.
With the exception of $\Rcmp$, each operator returns a function from desired precision $p \in \mathbb{Z}$ to an integer approximant.
Our library includes the following:
\begin{itemize}
  \item $\RofZ{}\ z$: This converts the constant integer $z$ to a real whose sequence is obtained by bit-shifting $z$.
  \item $\Radd{}\ r_1\ r_2$: Addition between $r_1$ and $r_2$. The first line adds the $(p+2)^{\textrm{nd}}$ approximants for $r_1$ and $r_2$ together, and the second line divides this sum by four (rounding to the nearest integer).
  \item $\Rneg{}\ r_1$: Negates a real number by negating each approximant.
  \item $\Rscal{}\ r\ z$: Uses a bit shift to approximate the real number $r/2^z$ for $z \in \mathbb{Z}$. CReal defines a more complicated procedure for general multiplication between real numbers, but scaling by powers of two will suffice for our purposes.
  \item $\RofU{}\ u$: Convert a partly sampled uniform deviate into a lazy real number. The $\GetBits{}$ function was defined in~\cref{fig:u-impl}.
\end{itemize}

Like $\cmpU$, the function $\Rcmp$ compares two real numbers by iteratively comparing approximants until they are sufficiently far apart to determine their inequality.\footnote{Note that the comparison function does not terminate when the arguments are equal. Unlike the comparison between two uniform deviates, it is possible to write programs where comparisons loop with probability greater than zero. }
In our Rocq development, we specify the correctness of these operators using the following \erisR{} predicate:
\begin{align*}
  \IsApprox\ (v : \Val)\ (x : \real)\ \eqdef \Exists I. I \sep \always {\forall (p : \mathbb{Z})\ldotp\ \hoare{I}{v\ p}{\Ret A. I \sep \ApproxTo(A, p, r)}}
\end{align*}

In other words, \IsApprox{} states that a \erisR{} function $v$ can be executed, using some set of resources $I$, in order to return an approximation of $r$ with any precision $p$.
Using this representation, it is straightforward to verify the correctness of the arithmetic operators in our library, such as the addition function.
\begin{align*}
\hoare
    {\IsApprox\ v_x\ x \sep \IsApprox\ v_y\ y}
    {\Radd\ v_x\ v_y}
    {v \ldotp \IsApprox\ v\ (x + y)}
\end{align*}
The correctness here depends principally on using the triangle inequality to relate the $(p+2)^{\textrm{nd}}$ approximants of $x$ and $y$ to the $p^\textrm{th}$ approximant of $x + y$.

We can also lift the lazily uniform samples of $\eunif{}$ into constructive reals.
\begin{align*}
\hoare
    {\isR\ u\ x}
    {\RofU\ u}
    {v \ldotp \IsApprox\ v\ x}
\end{align*}
In this proof, the $I$ in $\IsApprox$ is instantiated with $\isR\ u\ x$, which allows us to compute arbitrarily precise approximants for $x$ using $\GetBits{}$.

Finally, we will derive one last program for constructing values in $\mathbb{R}$ using these primitive operators.
The equation $\toR(b, z, r) \eqdef (-1)^b(z+r)$ will serve as our canonical mapping from boolean-integer-uniform triples $(b, z, r)$ to real numbers.
The program $\RofBZU$ in~\cref{fig:RofBZU} takes as input a boolean-integer-real triple, and returns the corresponding constructive real.
We will use this program in the verification of the Gaussian and Laplace samplers, and also our soundness theorem.

\begin{figure}
  \begin{align*}
    \RofBZU\ b\ z\ u & \eqdef {\begin{array}[t]{@{}l}
                          \Let \langv{abs} := \Radd\ (\RofZ\ z)\ (\RofU\ u) in  \\
                          \If b = 0 then \langv{abs} \Else \Rneg\ \langv{abs}
                          \end{array}}
  \end{align*}
  \caption{Conversion from boolean-integer-uniform triples into constructive reals.}
  \label{fig:RofBZU}
\end{figure}

\mbox{}

We have now seen how to establish and use a continuous analogue of the specification style that \citeauth{marionneau:samplers} previously demonstrated for verifying discrete samplers.
But what does this specification actually mean?
In the next section, we prove an adequacy theorem that connects this specification to the semantics of the program.

 \newcommand{\checker}{\langv{Checker}}

\section{Soundness}\label{section:soundness}

This section describes our soundness results for \erisR{}.
Before we can state the soundness results, we first need to define the semantics of the language we are considering formally.
Then, we prove that the basic adequacy theorem for Eris Hoare triples also holds for \erisR{}.
Finally, we prove an analogue of Marionneau et al.'s~\cite{marionneau:samplers} adequacy theorem for the correctness of continuous samplers.

\newcommand{\bind}{{\scriptstyle >\!>\!=}}
\newcommand{\munif}[1]{\distr_{\dunif{#1}}}
\newcommand{\dunif}[1]{\mathcal{U}(#1)}
\newcommand{\dempty}{\mathbf{0}}
\newcommand{\dirac}[1]{\delta_{#1}}

\subsection{\thelang{} Language}\label{subsection:language}

\begin{figure*}[ht]
  \begin{align*}
    \val \in \Val \eqdef{}
    & z \in \integer \ALT
      b \in \bool \ALT
      \TT \ALT
      \loc \in \Loc \ALT
      \lbl \in \Lbl \ALT
      \Rec \lvarF \lvar = \expr \ALT
      (\val,\valB) \ALT
      \Inl \val  \ALT
      \Inr \val \\
    \expr \in \Expr \eqdef{}  &
                                \val \ALT
                                \lvar \ALT
                                \expr_1~\expr_2 \ALT
  \expr_1 + \expr_2 \ALT
  \expr_1 - \expr_2 \ALT
  \expr_1 * \expr_2 \ALT
  \expr_1 \ll \expr_2 \ALT
  \expr_1 \gg \expr_2 \ALT
  \ldots \ALT \\
	  & \If \expr then \expr_1 \Else \expr_2 \ALT
    (\expr_1,\expr_2) \ALT
      \Fst \expr \ALT
      \Snd \expr \ALT
      \Inl(\expr) \ALT
      \Inr(\expr) \ALT \\
      & \Match \expr with \Inl \val~ => \expr_1 | \Inr \valB => \expr_2 end \ALT 
     \AllocN \expr_1~\expr_2 \ALT
      \deref \expr \ALT
      \expr_1 \gets \expr_2 \ALT \\
     & \Rand \expr \ALT
      \Rand \expr_1\, \expr_2 \ALT
      \AllocTape \expr
    \\
  \lctx \in \Ectx \eqdef{}  &
                               -
  \ALT \expr\,\lctx
  \ALT \lctx\,\val
  \ALT \AllocN\lctx
  \ALT \deref \lctx
  \ALT \expr \gets \lctx
  \ALT \lctx \gets \val
  \ALT \Rand \lctx \ALT \dots
    \\
	  t \in \Tape \eqdef{}& \{ (\tapebound, \tape) \mid \tapebound \in \mathbb{N} \wedge \tape \in \List{\mathbb{N}_{\leq \tapebound}} \} \\
    \state \in \State \eqdef{}& (\Loc \fpfn \Val) \times (\Lbl \fpfn \Tape)
        \qquad
    \cfg \in \Conf \eqdef{} \Expr \times \State \end{align*}
  \caption{Syntax of \thelang{}}\label{fig:syn}
\end{figure*}

\thelang{} is the same language as used in Eris.
The syntax of \thelang{} is depicted in \cref{fig:syn}.
The language includes primitives for higher-order programming such as higher-order references and function closures, as well as generic recursion using recursive let bindings.
\thelang's sole source of randomness is the $\Rand N$ command seen in the previous section.
Notably, \thelang{} does \emph{not} include primitives for sampling from continuous distributions.
Instead, we implement and verify computable versions of these algorithms.

Because the language only has discrete sampling as a primitive, we are able to give an operational semantics that does not require measure theory.
To account for the possibility of nontermination, we use the monad of discrete \emph{subdistributions}.

\begin{definition}[Subdistribution]
  A discrete subdistribution over a countable set $A$ is a function $\distr: A \to [0,1]$ such that $\sum_{a\in A} \distr(a) \le 1$,
  sometimes also called a \emph{probability mass function} or PMF.
  We let $\Distr{A}$ denote the set of all subdistributions over $A$.
  The discrete Giry monad~\cite{giry} is a monad on $\DDistr$ where the return $\dirac{x}$ is the Dirac distribution at $x$ and
  \begin{equation}
    (\distr \bind f)(b) \eqdef{} \sum_{a \in A} \distr(a) \cdot f(a)(b)
  \end{equation}
  Given a subset $\phi \subset A$ and $\mu \in \Distr{A}$, the probability of $\phi$ under $\mu$, denoted $\Pr_\mu[\phi]$, is
  given by
  \begin{equation*}
  \Pr_\mu[\phi] = \sum_{a\in \phi} [\mu(a)]
  \end{equation*}
\end{definition}

We give a small step operational semantics to \thelang{} in terms of subdistributions, following Eris.
Namely, we define the semantics for a single step of reduction by a monadic function on subdistributions $\stepdistr: \Conf \to \Distr{\Conf}$, where a configuration $\Conf$ is a pair of an expression $\Expr$ and a state $\State$.
For deterministic cases (such as arithmetic operations or writing to the store) we describe their posterior subdistribution using the monadic return.
Error cases such as ill-typed applications return the zero subdistribution $\dempty$.
The $\Rand\ N$ primitive draws samples from $\unifd{N}$, the uniform distribution over $\{0,\ldots,N\}$, and leaves the state $\sigma$ unchanged.
\begin{equation*}
  \stepdistr(\Rand\ N, \state) \eqdef{}
  \begin{cases}
    \unifd{N} \, \bind\, \lambda n\ldotp \dirac{(n, \sigma)} & \text{if } N \in \nat\\
    \dempty\ & \text{otherwise}
  \end{cases}
\end{equation*}

The semantics also models the presampling tapes we saw in \cref{sec:key-ideas}.
A tape consists of a \emph{label} $\alpha$, a \emph{bound} $N \in \mathbb{N}$, and a finite sequence of \emph{presamples} $\tape$.
When $\Rand$ includes the optional tape label parameter, it will deterministically pop a sample from the tape if the tape is non-empty, and otherwise draws a fresh random sample.
\begin{equation*}
  \begin{array}[t]{@{}l}
  \stepdistr(\Rand\ \alpha\ N, \state) \eqdef{} \\
  \quad \begin{cases}
     \dirac{(n,\state [\alpha\mapsto(N,\tape)])} & \text{if } N \in \nat \text{ and } \alpha \mapsto (N,n\cons\tape) \in \state \\
     \unifd{N} \, \bind\, \lambda n\ldotp \dirac{(n, \sigma)} & \text{if } N \in \nat \text{ and } \alpha \mapsto (N,n\cons\tape) \notin \state \\
			\dempty\ & \text{otherwise}
  \end{cases}
  \end{array}
\end{equation*}

Tape labels are dynamically allocated using the $\langkw{tape}$ primitive in a similar fashion to heap locations.
However, the \thelang{} language includes no language primitives for adding samples onto the tape.
Rather, samples are added to the end of the tape during a proof by applying the presampling rules we saw previously.
The soundness proof for the logic includes an \emph{erasure theorem}, showing that every program has the same semantics as one with all tape operations omitted.

The semantics of a program is the limit of finite executions.
Concretely, the subdistribution $\exec_n: \Conf \to \Distr{\Val}$ represents the subdistribution of values reached after $n$ steps of execution.
\begin{equation}
  \exec_n(\expr,\state) \eqdef{}
  \begin{cases}
    \dirac{e} & \text{if } \expr \in \Val \\
    \stepdistr(\expr,\state)\, \bind \, \exec_{n - 1} & \text{if } \expr \notin \Val \text{ and } n > 0\\
    \dempty & \text{otherwise}
  \end{cases}
\end{equation}
Any configuration that has not reached a value after $n$ steps does not contribute its probability to $\exec_n$.
The subdistribution of return values $\exec(\cfg)$ is defined at each point to be the limit of $\exec_n$ as $n \to \infty$.

\subsection{Soundness of the Logic}

We proved the following adequacy theorem of \erisR{}, which is the same as that of Eris:
\begin{sloppypar}
\begin{theorem}[Adequacy]\label{thm:adequacy}
  For any pure predicate on values $P$, if we can prove $\hoare{\upto{\epsilon}}{\expr}{\Ret \val. \prop~\val}$ in \erisR{},
  then for all states $\state$, we have $\Pr [\exec(\expr,\state)\notin \prop]\leq \epsilon$.
\end{theorem}
\end{sloppypar}
Recall that \erisR{} features two proof rules not found in the original Eris: thin-air credits and time receipts.
The thin-air credit rule was proved sound by \citeauth{coneris} for an extension to Eris for concurrent programs, and we have backported their proof technique to \erisR{}.
For time receipts, we exploit the fact that the existing proof of soundness for Eris is based on step-indexing: since the logic is partial, the proof works by explicitly considering executions of up to $n$ steps for each value of $n$.
Thus, all we need to do is track this number of steps remaining in the execution we are considering using a ghost resource, in order to model the $\stepsLeft$ assertion.
\joe{TODO: decide whether we want to elaborate on this, based on space or not. We could show the state interpretation/tweak to definition of wp, though it is pretty boring/simple, will depend upon the position of where this ends up going.}
\lars{a middle ground could be to write something like ``for the Iris experts, we remark that change the WP definition of Eris to track this ghost resource as part of the state interpretatio (and that we we have proved soundness of all the proof rules wrt. this updated definition of the WP)''.  This way we say that something has happened but also that it is pretty straightforward for Iris people, which I guess is what you mean by boring :-)}

\subsection{Adequacy for Continuous Samplers}

\newcommand{\cdfv}{H}
\newcommand{\pdfv}{h}

To formally capture the correctness of continuous samplers, we prove a version of the adequacy theorem tailored to continuous samplers returning lazy reals satisfying the $\IsApprox$ predicate.
Stating this is subtle: we want to specify what it means for a program to lazily return digits from a real number, but clearly this will depend on the \thelang{} representation of a real.
The proof of adequacy by \citeauth{marionneau:samplers} avoids this issue as their samplers all return simple, first-order datatypes such as $\mathbb{Z}$ or booleans, all of which have canonical representatives in \thelang{}.

\begin{figure}
  \begin{align*}
    \checker\ \langv{e}\ N\ D & \eqdef \Rcmp\ e\ (\Rscal\ (\RofZ\ N)\ D)\ 0
  \end{align*}
  \caption{Definition of the $\checker$ program.}
  \label{fig:checker}
\end{figure}

The key idea of our approach to this problem is to capture the behavior of a continuous sampler by analyzing the \emph{cumulative distribution function} (CDF) it converges to.
An absolutely continuous probability (sub-)distribution over the reals is characterized by a \emph{density function} $\pdfv \in \pcts(\mathbb{R})$, such that $\int_{-\infty}^\infty \pdfv(x) \,dx \leq 1$
and whose definite integrals $\int_a^b \pdfv(x) \,dx$ for $a \leq b$ correspond to the probability of sampling a value in the interval $[a,b]$.
Its corresponding CDF $\cdfv$ can be obtained by the equation $\cdfv(r) = \int_{-\infty}^r \pdfv(x) \,dx$.

We can observe the CDF of a \thelang{} real random sampler by linking it against the program $\checker$ in~\cref{fig:checker}.
The $\checker$ program takes in two arguments: the sampler $\langv{e}$ to be analyzed, and two integers $N$ and $D$.
Intuitively, $\checker$ samples from $\langv{e}$ and compares the sampled value to the dyadic rational number $N / 2^D$.
The $\checker$ program will diverge if the sampled value is exactly equal to $N/2^D$, otherwise it returns $-1$ or $1$ depending on whether $\langv{e}$ returns a constructive real less than or greater than $N / 2^D$, respectively.
Our adequacy theorem describes samplers in terms of their observable behavior when linked against this checker program, stated more precisely in \cref{thm:adequacy-sampler}.

\begin{theorem}[Partial Adequacy of Constructive Real Samplers]\label{thm:adequacy-sampler}
  Let $\pdfv \in \pcts(\mathbb{R})$ be a density function. If for all integrable $F \in \pcts(\mathbb{R})$ we can prove
  \begin{equation}
    \hoare{\upto{\int_{-\infty}^\infty F(x) \pdfv(x) \,dx}}{\langv{e}}{\Ret v. \Exists r. \upto{F\ r} \sep \IsApprox\ v\ r},
    \label{eqn:adequacy-pre}
  \end{equation}
  then for all states $\state$ and integers $B, C$ we have
  \begin{enumerate}
    \item $\Pr [\exec(\checker\ \langv{e}\ B\ C,\state)\neq 1]\leq\int_{-\infty}^{B/2^C} \pdfv(x)\,dx $
    \item $\Pr [\exec(\checker\ \langv{e}\ B\ C,\state)\neq -1]\leq\int_{B/2^C}^{\infty} \pdfv(x)\,dx $
  \end{enumerate}
\end{theorem}
This theorem relates the behavior of the $\checker{}$ program with the CDF of the target continuous distribution.
To understand this theorem, recall that $\int_{-\infty}^{B / 2^C} \pdfv(x) \,dx$ is the CDF of the distribution at $B /2^C$, \ie{} it is the probability that a value sampled from the distribution will be less than or equal to $B / 2^C$.
The first inequality says that the probability that $\checker$ terminates with a value other than $1$ is at most this CDF value.
The second inequality gives us an analogous result about the probability that the sampled value is greater than $B / 2^C$.

These inequalities give us a good sense of the approximate behavior of sampler programs when linked against $\checker{}$, however they do not completely characterize the CDF of $\langv{e}$ because \erisR{} is a partial logic.
It is possible that the sampling program fails to terminate with probability greater than zero.\footnote{This is only possible if the sampler program $\langv{e}$ fails to terminate. The $\checker$ program only fails to terminate when $\langv{e}$ returns the exact value $B/2^C$, which can be shown to happen with probability at most zero by instantiating~\cref{eqn:adequacy-pre} with the indicator function on $\{B/2^C\}$. }
By adding hypotheses to~\cref{thm:adequacy-sampler}, we obtain a stronger result.
\begin{theorem}[CDF Adequacy of Constructive Real Samplers]\label{thm:cdf-adequacy-sampler}
  Under the assumptions of~\cref{thm:adequacy-sampler}, if the checker program terminates almost-surely (\ie{} for all $B$, $C$, and $\sigma$ we have $\sum_{v \in \Val} \exec(\checker\ \langv{e}\ B\ C, \state)(v) = 1$) and $\pdfv$ is a probability density (\ie{} $\int_{-\infty}^{\infty} \pdfv(x)\, dx = 1$) then for all states $\state$ and integers $B, C$, the probability that $\checker{}\ \langv{e}\ B\ C$ returns $-1$ is equal to the CDF of $\pdfv$ at the point $B/2^C$:
  \begin{equation}
    \Pr [\exec(\checker\ \langv{e}\ B\ C,\state) = -1] = \int_{-\infty}^{B/2^C} \pdfv(x)\,dx.
    \label{eqn:cdf-adequacy}
  \end{equation}
\end{theorem}

Since the set of dyadic integers is dense in $\mathbb{R}$, characterizing the CDF at these points suffices to characterize the distribution of $\langv{e}$ completely (see \citeauth{billingsley1995}, theorem 14.1).

In our Rocq development, we have formally verified both of these adequacy statements.
The proof of~\cref{thm:cdf-adequacy-sampler} is a consequence of~\cref{thm:adequacy-sampler}, using the additional termination hypotheses in order to establish~\eqref{eqn:cdf-adequacy}.
For the programs we will present in sections~\cref{section:gauss} and~\cref{section:laplace}, we do not prove their almost-sure termination, though \citeauth{karney16} proves on paper that these algorithms terminate almost-surely.
Separate program logics such as Total Eris~\cite{eris} or Tachis~\cite{DBLP:journals/pacmpl/HaselwarterLMG024}, which cover the same source language, are capable of proving this fact.
For the remainder of this paper, we will focus on establishing the core specification~\eqref{eqn:adequacy-pre}, and leave the other side conditions in~\cref{thm:cdf-adequacy-sampler} to future work.

 \section{Verified Gaussian Sampler}\label{section:gauss}

In this section, we will verify a sampler for the continuous Gaussian distribution using \erisR{}.
We will use algorithms from~\cite{karney16}.
To the best of our knowledge, these algorithms have never been formally verified.
Karney's~\cite{karney16} sampler is split into a number of subroutines, and our proof follows the same modular structure by verifying each subroutine.

\subsection{Bernoulli Negative Half-Exponential}
\begin{figure}
  \begin{align*}
    \RealDecrTrial \ N \ x & \eqdef {\begin{array}[t]{@{}l}
                      \Let y := \eunif{} in \\
                      \If \cmpU\ y\  x < 0 then \RealDecrTrial\ (N+1)\ y \Else N
                      \end{array}}\\
	\HalfBernExpNeg \ \_     & \eqdef {\begin{array}[t]{@{}l}
                      \Let x := \eunif{} in \\
                      \If \RleHalf\ x then \RealDecrTrial\ 0\ x \Mod 2 = 1 \Else \True
                      \end{array}}
  \end{align*}
  \caption{
The decreasing uniform sequence trial ($\RealDecrTrial$) and the Bernoulli trial with parameter $e^{-1/2}$ ($\HalfBernExpNeg$).}
\label{fig:DecrTrial}
\end{figure}

The first component of Karney's~\cite{karney16} algorithm is a routine $\HalfBernExpNeg{}$ for drawing a sample from the $\textrm{Bernoulli}(e^{-1/2})$ distribution.
This in turn builds upon a routine $\RealDecrTrial{}$ which repeatedly draws a sequence of independent, uniform samples $r_1, r_2, r_3, \dots$ from $[0,1]$.
The routine stops when it draws an $r_{i+1}$ that is greater than $r_i$, at that point, it returns $i$, the length of the decreasing sequence it generated.

\paragraph*{Verifying the Decreasing Reals Trial}
The principle behind this algorithm is sometimes known as \emph{Von Neumann's Technique}~\cite{vonneumann51}.
Von Neumann's insight is that the probability that this decreasing sequence has length $n$ is $x^n/n!-x^{n+1}/(n+1)!$ --- two subsequent terms in the Taylor series of $e^{-x}$.

In general, $\RealDecrTrial{}\ N\ x$ is distributed over $\mathbb{N}$ as a version of $\RealDecrTrial{}\ 0\ x$ that has been shifted rightwards by $N$:
\begin{equation}
  \distrof{(\RealDecrTrial{}\ N\ x)}(n) \eqdef \iverbr{N \le n} \cdot \left(\frac{x^{n-N}}{(n-N)!}-\frac{x^{n-N+1}}{(n-N+1)!}\right)
  \label{eqn:realdecrtrial}
\end{equation}

Note that although this sampler draws continuous samples internally, what it \emph{returns} is a sample from a \emph{discrete} distribution over $\mathbb{N}$.
Thus, the specification we prove for it uses a series for computing the expected value of credits, instead of an integral.
Let $\bdd(\mathbb{N})$ denote the set of bounded functions $\mathbb{N} \to \real^{\ge 0}$.
We seek to prove the following specification for all $F \in \bdd(\mathbb{N})$, $N \in \mathbb{N}$, $l \in \Val$, and $x \in [0, 1]$:
\begin{equation}
\hoare
{\begin{array}[c]{@{}l}
  \isR\ l\ x \sep 0 \le x \le 1 \sep \\
  \upto{\sum_{n \in \mathbb{N}} \distrof{(\RealDecrTrial{}\ N\ x)}(n) \cdot F(n)}
\end{array}}
{\RealDecrTrial\ N\ x}
{n,
\begin{array}[c]{@{}l}
  \upto{F(n)} \sep \\
  \isR\ l\ x
\end{array}}
\label{eqn:realdecrtrialspec}
\end{equation}

At a high level we will prove this in a similar way to~\cref{subsection:emax}, using specification~\eqref{eqn:advcomp} at the call to $\eunif{}$ to redistribute the error credits based on the value of $y$.

The challenge in proving~\eqref{eqn:realdecrtrialspec} lies in correctly conditioning the credits
after each probabilistic sampling from $\eunif{}$.
We proceed by L\"ob induction, using~\rref{ht-rec}.
The execution of $\RealDecrTrial{}\ N\ x$ first evaluates the sampling statement $\eunif{}$, branches on its result, and either returns $N$ (requiring $\upto{F(N)}$ error credits for the postcondition) or makes a recursive call (requiring enough error credits to pay for the precondition of~\eqref{eqn:realdecrtrialspec} via~\rref{ht-rec}).
Schematically, we want to condition the credits after sampling from $\eunif{}$ as follows:
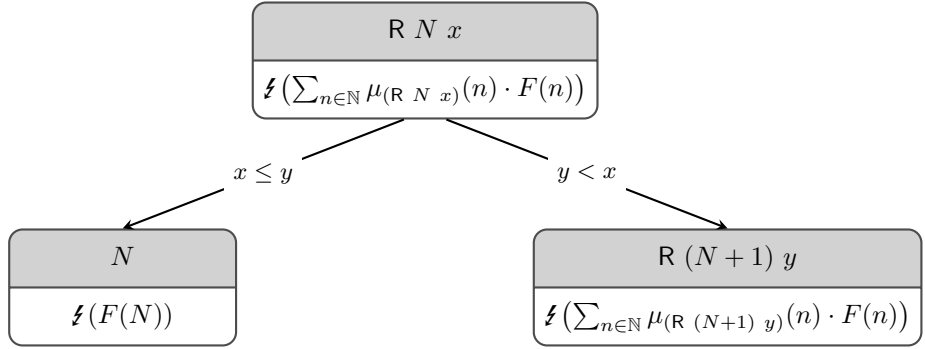
\begin{center}
\begin{tikzpicture}[
    node distance=-1cm,
detail/.style={font=\small\itshape},
splitnode/.style={
        rectangle split,
        rectangle split parts=2,
        rectangle split horizontal=false,  draw,
        rounded corners=5pt,
        minimum width=3cm,
        minimum height=1.4cm,
        text centered,
        rectangle split part fill={gray!35, white}, draw=black!70,
        thick,
        inner sep=4pt
    },
arr/.style={
        -{Stealth[length=3mm]},
        thick,
        draw=black!70
    },
lbl/.style={
        midway,
        fill=white,
        inner sep=2pt,
        font=\sffamily\small
    }
]
\node[splitnode] (A) at (0,0){
	\strut $\RealDecrTrial\ N\ x$
	\nodepart{two}
	\strut $\upto{\sum_{n \in \mathbb{N}} \distrof{(\RealDecrTrial{}\ N\ x)}(n) \cdot F(n)}$
	};
\node[splitnode] (B) at (-4,-3){
	\strut $N$
	\nodepart{two}
	\strut $\upto{F(N)}$
	};
\node[splitnode] (C) at (4,-3){
	\strut $\RealDecrTrial\ (N+1)\ y$
	\nodepart{two}
	\strut $\upto{\sum_{n \in \mathbb{N}} \distrof{(\RealDecrTrial{}\ (N+1)\ y)}(n) \cdot F(n)}$
	};
\draw[->, thick, >=stealth] ([xshift=-8pt]A.south) -- (B.north)
    node[midway, fill=white, detail] {$x \le y$};
\draw[->, thick, >=stealth] ([xshift=8pt]A.south) -- (C.north)
    node[midway, fill=white, detail] {$y < x$};
\end{tikzpicture}
\end{center}
With this in mind, we choose the following function $g$ to condition the error credits:
\begin{equation}
    g(F, N, x, y) \eqdef 
	\iverbr{x \le y} \cdot F(N) +
	\iverbr{y \le x} \left(\sum_{m \in \mathbb{N}} \distrof{(\RealDecrTrial{}\ (N+1)\ y)}(m) \cdot F(m)\right)
  \label{eqn:gdef}
\end{equation}
After sampling from $\eunif{}$ we obtain a lazy real $y$ and $\upto{g(F, N, x, y)}$ error credits.
However, note that at this point we have no further information about $y$, which we need
to evaluate the comparison $\cmpU\ y\  x$.
Therefore, we perform a case distinction on $(x \le y) \vee (y \le x)$, which can be
used to cancel out the corresponding Iverson bracket in~\eqref{eqn:gdef} and keep executing
the program symbolically\footnote{The fact that the Iverson brackets are overlapping at the point $y=x$ is a matter of convenience. }. In the first case, we will own $\upto{F(N)}$, and, by symbolic execution, the program will
evaluate to the $\langkw{else}$ branch and
return $N$, so the postcondition is satisfied. In the second case, we will own $\upto{\sum_{m \in \mathbb{N}} \distrof{(\RealDecrTrial{}\ (N+1)\ y)}(m) \cdot F(m)}$, and we will reach the recursive call to $\RealDecrTrial\ (N+1)\ y$,
at which point we
can use the inductive hypothesis since we have the correct amount of credits available.
It remains to show that the expected value of $g$ is equal to our initial quantity of starting credits.
\begin{equation}
  \sum_{n \in \mathbb{N}} \distrof{(\RealDecrTrial{}\ N\ x)}(n) \cdot F(n) = \int_0^1 g(F, N, x, y) \,dy
  \label{eqn:gexpectedvalue}
\end{equation}
The proof of this fact amounts to an application of Fubini's theorem in a similar style to~\cref{subsection:emax}.
Here, the boundedness of $F$ ensures absolute and uniform convergence of the limits in~\eqref{eqn:gexpectedvalue}, allowing their exchange.

\paragraph*{Bernoulli Negative Half-Exponential}
Finally, our specification for $\RealDecrTrial$ can be used to verify the correctness of the half-exponential Bernoulli sampler.\footnote{In our implementation of $\HalfBernExpNeg$, we must handle the first iterate of this process separately, since $\cmpU$ can only compare two uniform deviates against each other, and so we need to use a specialized program $\RleHalf$ to compare a uniform deviate against the constant $1/2$. We choose to do it this way instead of lifting to constructive real library in order to mimic more closely Karney's implementation. }
We seek to prove that for any $F : \{\True, \False\} \to \real^{\ge 0}$,
\begin{equation}
\hoare
{\upto{e^{-1/2} F(\True) + (1 - e^{-1/2}) F(\False)}}
{\HalfBernExpNeg}
{b, \upto{F(b)}}
\label{eqn:hbne_spec}
\end{equation}
The uniform sample in $\HalfBernExpNeg$ is handled by the credit distribution
\begin{equation}
  \bigg[\frac{1}{2} < x \bigg] F(\True) +
  \bigg[x \le \frac{1}{2}\bigg] \left(\sum_{m \in \mathbb{N}} \distrof{(\RealDecrTrial{}\ 0\ x)}(m) \cdot F(m \% 2 = 1)\right)
  \label{eqn:gdef2}
\end{equation}
In the case that $1/2 < x$ we gain the correct amount of credits to return $F(\True)$, and if not, we have enough to apply the specification for $\RealDecrTrial$.
The function $\RleHalf$ is a version of $\cmpU$ which compares a lazy real sample against the constant value $1/2$, and is verified in a similar manner.
In our development, we split the series in expression~\eqref{eqn:gdef2} into even and odd terms, at which point we can apply Von-Neumann's trick.
\begin{equation*}
\sum_{n \in \mathbb{N}} \frac{(1/2)^{2n}}{(2n)!}-\frac{(1/2)^{2n+1}}{(2n+1)!}
= \sum_{n \in \mathbb{N}} \frac{(-1/2)^n}{n!}
= e^{-1/2}
\end{equation*}

\subsection{Integer Gaussian Part}\label{subsection:gaussint}

\begin{figure}
  \begin{align*}
    \GeoTrial\ e \ N & \eqdef \ \If e \ \Unit then \GeoTrial \ e \ (N+1) \Else N \\
	\IterTrial\ e \ N & \eqdef \ (N = 0) \Or (e\ \Unit \And \, \IterTrial\ e \ (n-1)) \\
    \GaussInt \ \_ & \eqdef {\begin{array}[t]{@{}l}
                               \Let k := \GeoTrial\ \HalfBernExpNeg \ 0 in \\
                               \If \IterTrial\ (\HalfBernExpNeg\ (k*(k-1))) then k \Else \GaussInt\ \Unit
                               \end{array}}
  \end{align*}
  \caption{The integer part of the Gaussian sampler.}
  \label{fig:GaussInt}
\end{figure}

The next component of the algorithm is a routine to sample the \emph{integer} part of the Gaussian sample, as shown in \cref{fig:GaussInt}.
Specifically, the routine $\GaussInt$ draws from the distribution over the nonnegative integers with PMF
\begin{equation}
  \distrof{\GaussInt}(k) \eqdef e^{-k^2/2} / \normof{\GaussInt} \quad\quad\quad \normof{\GaussInt} \eqdef \sum_{k=0}^\infty e^{-k^2/2}
  \label{eqn:gaussintpmf}
\end{equation}

The implementation uses helper functions $\GeoTrial$ (a geometric trial, parameterized by a function $e$ that generates Bernoulli samples) and $\IterTrial$ (which draws $N$ Bernoulli samples using argument $e$ and returns $\True$ when all samples are $\True$).
Our proofs of these are exactly analgous to those described by \citeauth{marionneau:samplers} and we will not replicate them here, however we highlight that it is important that this program is higher-order, as the Gaussian sampler dynamically chooses the parameters to $\GeoTrial$ and $\IterTrial$.
Using these combinators we can implement $\GaussInt$, Karney's sampler for the integer part of the Gaussian distribution.\footnote{This sampler corresponds to steps \texttt{N1} and \texttt{N2} from Karney's paper.}

The program itself is a simple rejection sampling loop, and involves no real numbers beyond those used internally by $\HalfBernExpNeg$.
Proving that $\GaussInt$ is distributed as $\distrof{\GaussInt}$ amounts to applying the specifications we have developed so far.
At a high level, we begin with $\upto{V}$ credits where $V \eqdef \sum_{k=0}^\infty \distrof{\GaussInt}(k) \cdot F(k)$ for some $F \in \bdd(\mathbb{N})$.
Like before, the proof proceeds by L\"ob induction, conditioning the credits on the results of $\GeoTrial$ and $\IterTrial$ by $g$ and $h$ respectively:
\begin{align*}
h(k, b) & \eqdef \iverbr{b} \cdot F(k) + \iverbr{\neg b} \cdot V \\
  g(k) & \eqdef e^{-k(k-1)/2} h(k, \True) + (1-e^{-k(k-1)/2}) h(k, \False)
\end{align*}

The fact that their expected value does not exceed the initial amount of error credits follows from Fubini's theorem.

\subsection{Gaussian Sampler}

\begin{figure}
  \begin{align*}
    \IntChoice\ M & \eqdef {\begin{array}[t]{@{}l}
                       \Let m := \Rand M in \\
                       \If m = 0 then -1 \Else \If m = 1 then 0 \Else 1
                       \end{array}} \\
    \FractionReject \ k\ x & \eqdef {\begin{array}[t]{@{}l}
                                   \Let f := \IntChoice\ k in \\
                                   \Let r := \eunif{} in \\
                                   (f = 0) \Or (f = -1 \And \cmpU\ x\ r < 0 )
                                   \end{array}} \\
    \Selector \ k\ x \ y\ N  & \eqdef {\begin{array}[t]{@{}l}
                                      \Let z := \eunif{} in \\
                                      \If (\cmpU \ y\  z < 0 \Or \FractionReject\ k\ x) then N \Else (\Selector\ k\ x\ z\ (N + 1))
                                      \end{array}} \\
    \SelectorZ \ k\ x & \eqdef{\begin{array}[t]{@{}l}
                                      \Let z := \eunif{} in \\
                                      \If (\cmpU\ x \ z < 0 \Or \FractionReject\ k\ x) then 0 \Else (\Selector\ k\ x\ z\ 1)
                                      \end{array}} \\
    \GaussReject \ k\ x & \eqdef \ (\SelectorZ \ k \ x) \Mod 2 = 0 \\
    \Gauss \ \_ & \eqdef {\begin{array}[t]{@{}l}
                       \Let k := \GaussInt\ \Unit in \\
                       \Let x := \eunif{} in \\
                       \If \IterTrial\ (\GaussReject\ k\ x) \ (k+1) then (k, x) \Else \RealDecrTrial\ \Gauss\Unit
                       \end{array}} \\
    \GaussSymm\ \_ & \eqdef  {\begin{array}[t]{@{}l}
                              \Let (z, u) := \Gauss\ \Unit in \\
                              \Let b := \Rand \ 1 in \\
                              \RofBZU\ b\ z\ u
                              \end{array}}
  \end{align*}
  \caption{The Gaussian sampler ($\GaussSymm$).}
  \label{fig:Gauss}
\end{figure}

A \thelang{} implementation of Karney's sampler for the Gaussian distibution is given in $\Gauss{}$ in~\cref{fig:Gauss}.
More precisely, the $\Gauss$ sampler draws from the \emph{half} or \emph{one-sided} Gaussian distribution over $[0, \infty)$, represented as a pair containing a nonnegative integer and a lazy uniform real.
The full Gaussian sampler $\GaussSymm$  transforms a sample from $\Gauss$ into the full normal distribution over $(-\infty, \infty)$ in $\GaussSymm$ by choosing a sign uniformly at random, and applying $\RofBZU$ to combine our samples into our constructive real library from~\cref{sec:keyieas:creal}.

The implementation of $\Gauss{}$ involes a complicated rejection step $\GaussReject{}$, which (like Karney) we split into a number of steps:
\begin{itemize}
  \item $\IntChoice\ M$: returns $-1$, $0$, and $1$ with probabilities $1/M$, $1/M$, and $(M-2)/M$ respectively, when $1 \le M$.
  \item $\FractionReject\ k\ x$: A bernoulli trial with parameter $(1-(2k+x)/(2k+2))$.
  \item $\SelectorZ\ k\ x$: A $\Distr{\mathbb{N}}$ sampler with PMF
    \[
      \frac{x^n}{n!} {\left(\frac{2k+x}{2k+2}\right)}^{n} - \frac{x^{n+1}}{(n+1)} {\left(\frac{2k+x}{2k+2}\right)}^{n+1}
    \]
    $\Selector$ is a sampling loop for this program, written in tail-recursive form. This program is similar to the loop in $\RealDecrTrial{}$, but with additional \emph{thinning} introduced by $\FractionReject{}$.
  \item $\GaussReject \ k\ x$: A $e^{-x(2k+x)/(2k+2)}$ Bernoulli trial, based on Von-Neumann's technique.
\end{itemize}

Aside from the spike in complexity, the techniques we have developed so far are capable of verifying this composition of samplers with no substantial modifications.
In our Rocq development we prove specifications for each helper function in the style of~\eqref{eqn:GenExpectation}, quantifying over functions in $\bdd(\{-1, 0, 1\})$, $\bdd(\{\True, \False\})$, or $\bdd(\nat)$, which are used modularly to condition error credits in their clients.
Each helper function is a discrete sampler, though some of them (like $\Selector$) make internal use of real numbers to sample discrete values with the correct probabilities.
\Appref{appendix:credits} contains the credit distribution functions for each of these helpers, as well as the proof that their expected value does not exceed the initial supply of error credits.
The most challenging aspect of our approach is determining a closed form for the distribution of each helper function, which is necessary to state specifications in the form of~\eqref{eqn:GenExpectation}.

With $\GaussReject{}$ in hand, we can verify the correctness of the one-sided Gaussian sampler $\Gauss{}$.
Because the return value for this program is a pair of a (nonnegative) integer and a lazy real, our specification expresses the density of $\Gauss{}$ as
\begin{equation}
  \distrof{\Gauss}(n, x) \eqdef e^{-(n+x)^2/2} / \normof{\Gauss} \quad\quad \normof{\Gauss} \eqdef \int_0^1 \sum_{k=0}^\infty e^{-(k+x)^2/2} \,dx
  \label{eqn:gausspmf}
\end{equation}

Finally, we define the symmetric normal distribution $\GaussSymm$ by lifting our Gaussian samples to a CReal, choosing the sign uniformly at random.
For any bounded, piecewise continuous $F : \mathbb{R} \to \mathbb{R}^{\ge 0}$ we can show that
\begin{equation}
  \hoare{\upto{\int_{-\infty}^\infty \frac{e^{-x^2/2}}{2\normof{\Gauss}} \cdot F(x) \,dx}}{\GaussSymm\ \Unit}{\Ret v. \Exists r. \IsApprox\ v\ r \sep \upto{F(r)}}.
  \label{eqn:symmgaussspec}
\end{equation}
To prove this, we instantiate the specification for $\Gauss$ with the function $G(n, x) = F(-(n+x)) + F(n+x)$, and then use the final $\Rand\ 1$ step to distribute that credit to the positive and negative cases appropriately.
Our specification for $\RofBZU$ performs the final task of lifting this result to a CReal.
The specification \eqref{eqn:symmgaussspec} is now in a form for which we can apply the adequacy theorems, and this completes our verification.

\section{Verified Laplace Sampler}\label{section:laplace}
In this section, we verify a sampler for the continuous Laplace distribution $\laplacepmf{\varepsilon, \mu}$, which has important applications in differentially private algorithms.
Its density function is given by:
\begin{equation}
  \laplacepmf{\varepsilon, \mu}(x) = \frac{\varepsilon}{2} \cdot e^{-\varepsilon \cdot |x - \mu|}
  \label{eqn:laplacedef}
\end{equation}
Our starting point is a sampler for the negative exponential distribution~(\cref{sec:laplace:neg-exp}).
Samples from this distribution can be transformed into Laplace samples by applying appropriate arithmetic operations, which we implement using our verified library for constructive real arithmetic~(\cref{sec:keyieas:creal}).
Finally, we prove a tail bound on the size of Laplace samples, which has applications for proving \emph{accuracy} bounds of differentially private algorithms~(\cref{sec:laplace:accuracy}).

\subsection{Laplace Sampling}\label{sec:laplace:neg-exp}

\begin{figure}
  \begin{align*}
    \NegExp \ N & \eqdef {\begin{array}[t]{@{}l}
                       \Let x := \eunif{} in \\
                       \Let y := \RealDecrTrial\ 0\ x in \\
                       \If y \Mod 2 = 0 then (x, N) \Else \NegExp\ (N+1)
                       \end{array}}
  \end{align*}
  \caption{The Negative Exponential sampler ($\NegExp$).}
  \label{fig:NegExp}
\end{figure}

The Laplace sampler uses the sampler $\NegExp$ for the negative exponential distribution shown in \cref{fig:NegExp}.
Karney~\cite{karney16} calls this \emph{Algorithm V}.
Similarly to the one-sided Gaussian sampler, $\NegExp$ returns pairs in $\nat \times [0,1]$.
In this case, the probability of returning $(z, x)$ is $\distrof{\NegExp} \eqdef e^{-(z+x)}$.
The proof of this fact is similar enough to the proof of the Gaussian sampler in~\cref{section:gauss} that we will omit it and direct interested readers to the appendix or Rocq development.

We must make three modifications to the exponential mechanism in order to turn it into a sampler for the Laplace distribution:
\begin{enumerate}
  \item Extend the distribution by symmetry to all of $\mathbb{R}$,
  \item Multiply the result by a scaling factor $\varepsilon$, and
  \item Shift the distribution by the mean $\mu$.
\end{enumerate}

We can implement all three of these steps using our constructive reals library as described in~\cref{sec:keyieas:creal}.
\Cref{fig:Laplace} depicts the implementation of our Laplace sampler.
First, after taking a sample from $\NegExp$, one can decide the sign of the result using a $\Rand 1$ sample.
The boolean-integer-uniform triple is lifted into a single constructive real value using the $\RofBZU$ function which, like the Gaussian, gives us a sample from the symmetric version of the negative exponential distribution.
Now we are able to perform arithmetic on this sample: we can scale and shift the returned value using our constructive reals library.
\begin{figure}
  \begin{align*}
    \LaplaceSample \ \epsilon \ \mu & \eqdef {\begin{array}[t]{@{}l}
                                            \Let (z, u) := \NegExp\ 0 in \\
                                            \Let \langv{sgn} := \Rand \ 1 in \\
                                            \Radd \ \mu \ (\Rscal\ \epsilon\ (\RofBZU \ (\langv{sgn}, z, u)))
                                            \end{array}}
  \end{align*}
  \caption{The Laplace sampler ($\LaplaceSample$).}
  \label{fig:Laplace}
\end{figure}

After obtaining the sample from the negative exponential function and symmetrizing it like we did in~\cref{section:gauss}, the remaining operations are all deterministic.
Putting everything together, we can prove that for any $F \in \pcts(\real)$,
\begin{equation}
  \hoare{
    \upto{\int_{-\infty}^\infty F(x) \laplacepmf{2^\varepsilon, \mu} \,dx}
    \sep \IsApprox\ v_\mu\ \mu\ I_{\mu}
  }
  {\LaplaceSample\ \varepsilon \ v_\mu}
  {\Ret v. \Exists I_{L}, r.
    \begin{array}[c]{@{}l}
      \upto{F\ r} \sep I_{L} \sep \\
      \IsApprox\ v\ r\ (I_{\mu} \sep I_{L})
    \end{array}
   }
  \label{eqn:lapspec}
\end{equation}

\subsection{Accuracy Bound}\label{sec:laplace:accuracy}
As mentioned previously, the continuous Laplace distribution is widely used in differential privacy applications.
With differential privacy, random noise is added to the results of computations to avoid leaking private information about the data used in the computation.
Adding more noise gives stronger privacy guarantees, but adding too much noise can make the computed value less useful.
Thus, an important consideration in analyzing differentially private algorithms is to consider the \emph{utility} or \emph{accuracy} of the algorithm.
For example, in defining the \emph{Report Noisy Max} algorithm, \citeauth{dwork14} establish a utility bound on a differentially private query program based on the following property of the Laplace distribution:
 \begin{equation}
   \mathbb{P}\left[\left|\laplacepmf{\varepsilon, \mu} - \mu \right| > \frac{\log\left(1/\beta\right)}{\varepsilon}\right] \le \beta
   \label{eqn:accbound}
 \end{equation}
\citeauth{ub} incorporated a similar bound for the \emph{discrete} Laplace distribution as a rule in aHL, a program logic for reasoning about error bounds, where sampling from the discrete Laplace was added as a primitive to the language.

Instead of adding this bound as a primitive rule in our logic, our specification for the continuous Laplace sampler allows us to derive this bound for our sampler implementation in a straightforward way.
First, we instantiate the specification~\eqref{eqn:lapspec} with $F$ being the indicator function for the set
\[
S \eqdef \bigg(-\infty, \mu-\frac{\log(1/\beta)}{2^\varepsilon}\bigg] \cup \bigg[\mu+\frac{\log(1/\beta)}{2^\varepsilon}, \infty\bigg)
\]
Then, computing the improper integral in~\eqref{eqn:lapspec}, we can use the fact that $F(x) = 1$ for $x \in S$ together with the \emph{spending} rule to prove the following \erisR{} equivalent of the approximate specification~\eqref{eqn:accbound}:
\begin{equation*}
  \hoare{
    \upto{\beta} \sep \IsApprox\ v_\mu\ \mu\ I_{\mu}
  }
  {\LaplaceSample\ \varepsilon \ v_\mu}
  {\Ret f. \Exists I_{L}, r.
    \begin{array}[c]{@{}l}
      (|r-\mu| < \frac{\log\left(1/\beta\right)}{2\varepsilon}) \sep I_{L} \sep \\
      \IsApprox\ v\ r\ (I_{\mu} \sep I_{L})
    \end{array}
   }
\end{equation*}

\section{Related Work}

We have already alluded to the extensive literature on different representations of computable reals and operations on them.
The implementation we have verified is a simplified adaptation of the CReal library~\cite{creal} which is based on the work of \citeauth{menissiermorain:hal-02545650}.
The actual CReal library implementation uses additional mutable state to cache approximations of numbers for efficiency.
Since our language includes mutable state, it should be possible to generalize our verification to handle this as well.

SampCert~\cite{sampcert} verifies a number of discrete samplers in the Lean theorem prover.
Unlike \erisR{}, which supports the verification of programs that use higher-order random functions and state (features known to be challenging to reason about denotationally), SampCert's object language is restricted to first-order programs.
Hence, it cannot support samplers based on lazy reals.
The SampCert development includes a verified implementation of the discrete Gaussian sampler, a two-sided version of our sampler $\GaussInt{}$ from~\cref{subsection:gaussint} which can have variance $\sigma \in \mathbb{Q}$ and mean $\mu \in \mathbb{Z}$.

Zar~\cite{DBLP:journals/pacmpl/Bagnall0023} is a formally verified compiler for discrete probabilistic programs that generates executable samplers.
Its proof shows that the generated code produces samples with the correct probability distribution.

Although we have focused on two exact samplers for continuous distributions, a range of algorithms exist for different distributions and stochastic processes.
\citeauth{huber2016perfect} and \citeauth{occil} provide extensive overviews of these algorithms.
A related line of work explores what probability distributions are computable, \ie in the sense that one can computably approximate the value of the measure or density defining a distribution~\cite{FreerR12, AckermanFR19, HOYRUP2009830, DBLP:journals/iandc/Edalat09, DBLP:conf/csl/BilokonE14}.

As alluded to earlier, the semantics of languages that have sampling from continuous probability distributions as a primitive poses some technical challenges, particularly when combined with other semantically rich language features, such as higher-order functions and general recursion.
A number of approaches to the denotational semantics of higher-order probabilistic programs that feature continuous sampling have been proposed~\cite{DBLP:conf/lics/HeunenKSY17, DBLP:journals/pacmpl/EhrhardPT18, DBLP:journals/pacmpl/DahlqvistK20, DBLP:journals/pacmpl/VakarKS19, DBLP:conf/lics/HuotLMS23}.
Others have explored operational semantics for such languages~\cite{DBLP:conf/icfp/BorgstromLGS16} and logical relations techniques for proving equivalences~\cite{DBLP:journals/pacmpl/WandCGC18}.
Because the language we consider only has discrete probabilistic sampling as a primitive, and uses this to implement computable versions of continuous sampling, we avoid the semantic challenges of continuous distributions, while meanwhile still handling other features such as mutable higher-order state.
\citeauth{DBLP:conf/esop/HuangM16} used computable distributions as the basis for a denotational semantics of a probabilistic programming language, and gave a sampler implementation for this approach.

\citeauth{dash:lazy} develop a monadic probabilistic programming language with support for laziness, and show how to use it to express various stochastic processes, such as Poisson and Gaussian processes.
Since these processes are an infinite collection of random variables, they cannot be sampled in their entirety.
Instead, a key idea is to lazily sample parts of a process and then cache the sampled values, similar to how the $\eunif$ sampler we considered lazily samples and stores the bits of a uniform $[0,1]$ sample.
Their language supports a form of conditioning, and can be used for Bayesian modeling with a Monte Carlo inference algorithm.
They give a denotational semantics in terms of quasi-Borel spaces and an implementation called LazyPPL in Haskell.
LazyPPL assumes the existence of primitives for sampling directly from continuous distributions such as the Gaussian, which are implemented in practice using floats.
In contrast, our focus has been on proving the correctness of computable samplers for these continuous distributions, and we have used an operational semantics.

Recently, a number of program logics~\cite{lilac, basl, Sato16, SatoABGGH19, HirataMS22} have been developed for reasoning about programs that sample from continuous probability distributions.
These logics all treat sampling from continuous distributions and exact real operations as primitives, as opposed to our approach of verifying computable implementations of these operations.
Moreover, these logics do not support mutable state or dynamically allocated memory.
In contrast, \erisR{} inherits the Iris framework's support for reasoning about state and pointers.
On the other hand, some of these logics support language constructs for conditioning on observable events, as needed for Bayesian probabilistic modeling, which \erisR{} does not handle.

The proof outlined in \cref{sec:keyideas:continuous} for generalizing Eris's discrete error credit rules into continuous versions exploits the fact that the Riemann integral can be approximated by Riemann sums. 
\citeauth{BatzKRW25} have also used Riemann sums to approximate integrals in verifying probabilistic programs.
They observe that when using a pre-expectation calculus for a program with continuous samples, the standard approach gives rise to terms involving integrals, which poses a challenge for automated verification.
They show that by instead replacing these integrals with approximations by Riemann sums, one obtains a version that is more amenable to automation, while still yielding sound bounds on the original problem.
\citeauth{BeutnerOZ22} similarly use Riemann sums in bounding properties of probabilistic programs.

Although it is common to use floating-point approximations of continuous distributions, \citeauth{GargHBM24} instead consider using \emph{fixed-point} approximations in a probabilistic programming language for Bayesian inference.
They show that it is then possible to apply exact techniques for discrete Bayesian inference to this approximation, while soundly controlling the errors introduced by discretization.

 \section{Conclusion}

Exact sampling algorithms for continuous distributions involve a combination of features that are challenging to reason about.
\erisR{} provides a way to verify these algorithms using infinite presampling tapes, which are derived by extending Eris with time receipts.

\subsection{Future Work}

There are other logics related to Eris that also feature presampling tapes and mechanisms that behave similarly to error credits, such as Approxis~\cite{DBLP:journals/pacmpl/HaselwarterLAGTB25}, which allows for relational reasoning, and Tachis~\cite{DBLP:journals/pacmpl/HaselwarterLMG024}, for bounding expected costs.
It would be interesting to also extend these logics with time receipts and derive infinite presampling tapes to be able to apply them to programs that use exact samplers from continuous distributions.

Another direction for future work is to address the restrictions associated to Riemann integration.
One concrete limitation of the current approach is that our credit distribution functions are required to be piecewise continuous, and therefore cannot express credit distributions which have an uncountable number of discontinuities.
Integrability notwithstanding, one might imagine using the indicator function on the rationals as a credit distribution function to prove that the uniform sampler avoids all rational points.
In its current form proving this in \erisR{} requires one to apply~\cref{thm:cdf-adequacy-sampler}, thereby exiting the program logic and giving up the ability to use this fact as part of a larger \erisR{} proof.
Generalizations to more permissive integrals such as the Lebesgue or Henstock–Kurzweil integral may alleviate these side conditions.

\bibliography{refs}

\ifbool{fullversion}{
\pagebreak
\appendix
\onecolumn
\section{Appendix}\label{sec:appendix}

\subsection{Credit Conditioning}\label{appendix:credits}

Here we describe the credit conditioning functions used by our proofs, and show a proof that they are expectation-preserving.
Many of the derivations involve an exchange of limits, for which we use Fubini's theorem.
Formally justifying the existence and exchange of these iterated limits typically amounts to proving that their arguments converge uniformly, or are uniformly continuous.
For more details we refer the reader to our Rocq development, where all of the equalities in this section are formally verified.

\subsubsection{Decreasing Reals Trial}

\begin{itemize}
  \item For the uniform deviate sample from $\eunif{}$, use
    \begin{align*}
      g(F, N, x, y) \eqdef \iverbr{y \le x} \left(\sum_{m \in \mathbb{N}} \distrof{(\RealDecrTrial{}\ (N+1)\ y)}(m) \cdot F(m)\right) + \iverbr{y \ge x} \cdot F(N).
    \end{align*}
\end{itemize}

\paragraph{Expectation Preservation}
\begin{small}
\begin{align*}
    & \int_0^1 g(F, N, x, y) \,dy \\
  = & \int_0^1 \iverbr{y \le x} \left(\sum_{m \in \mathbb{N}} \distrof{(\RealDecrTrial{}\ (N+1)\ y)}(m) \cdot F(m)\right) + \iverbr{y \ge x} \cdot F(N) \,dy \\
  = & \int_0^1 \iverbr{y \le x} \left(\sum_{m \in \mathbb{N}} \distrof{(\RealDecrTrial{}\ (N+1)\ y)}(m) \cdot F(m)\right) \,dy + \int_0^1 \iverbr{y \ge x} \cdot F(N) \,dy \\
  = & \int_0^1 \iverbr{y \le x} \left(\sum_{m \in \mathbb{N}} \iverbr{N + 1 \le m} \cdot \left(\frac{y^{m-N-1}}{(m-N-1)!}-\frac{y^{m-N}}{(m-N)!}\right) \cdot F(m)\right) \,dy  \\
    & \quad + \int_0^1 \iverbr{y \ge x} \cdot F(N) \,dy \\
  = & \sum_{m \in \mathbb{N}} \left[ \int_0^1 \iverbr{y \le x} \left(\iverbr{N + 1 \le m} \cdot \left(\frac{y^{m-N-1}}{(m-N-1)!}-\frac{y^{m-N}}{(m-N)!}\right) \right) \,dy \cdot F(m)  \right] \\
    & \quad + \int_0^1 \iverbr{y \ge x} \cdot F(N) \,dy \\
  = & \sum_{m=N+1}^\infty \left[ \int_0^x \left(\frac{y^{m-N-1}}{(m-N-1)!}-\frac{y^{m-N}}{(m-N)!}\right) \,dy \cdot F(m) \right] + \int_x^1 F(N) \,dy \\
  = & \sum_{m=N+1}^\infty \left[ \left(\frac{x^{m-N}}{(m-N)!}-\frac{x^{m-N+1}}{(m-N+1)!}\right) \cdot F(m) \right] + (1-x) F(N) \\
  = & \sum_{m=N}^\infty \left[ \left(\frac{x^{m-N}}{(m-N)!}-\frac{x^{m-N+1}}{(m-N+1)!}\right) \cdot F(m) \right] \\
  = & \sum_{n \in \mathbb{N}} \distrof{(\RealDecrTrial{}\ N\ x)}(n) \cdot F(n)
\end{align*}
\end{small}

\subsubsection{Bernoulli Negative Half Exponential}

\begin{itemize}
  \item For the uniform deviate sample from $\eunif{}$, use
    \begin{align*}
      g(F, x) \eqdef \iverbr{1/2 < x} \cdot F(\True) + \iverbr{x \le 1/2} \left(\sum_{m \in \mathbb{N}} \distrof{(\RealDecrTrial{}\ 0\ x)}(m) \cdot h(F, m)\right)
    \end{align*}
  \item For the integer sample from $\RealDecrTrial{}$ use
    \begin{align*}
      h(F, n) \eqdef F(n \% 2 = 1)
    \end{align*}
\end{itemize}

\paragraph{Expectation Preservation}
\begin{small}
\begin{align*}
      & \int_0^1 g(F, x) \, dx \\
    = & \int_0^1 \iverbr{1/2 < x} \cdot F(\True) + \iverbr{x \le 1/2} \left(\sum_{m \in \mathbb{N}} \distrof{(\RealDecrTrial{}\ 0\ x)}(m) \cdot F(m \% 2 = 1) \right) \, dx \\
    = & \frac{F(\True)}{2} + \int_0^1 \iverbr{x \le 1/2} \left(\sum_{m \in \mathbb{N}} \distrof{(\RealDecrTrial{}\ 0\ x)}(m) \cdot F(m \% 2 = 1) \right) \, dx \\
    = & \frac{F(\True)}{2} + \sum_{m \in \mathbb{N}} \left[ \int_0^1 \iverbr{x \le 1/2} \left(\distrof{(\RealDecrTrial{}\ 0\ x)}(m) \right) \, dx \cdot F(m \% 2 = 1) \right] \\
    = & \frac{F(\True)}{2} + \sum_{m \in \mathbb{N}} \left[ \int_0^{1/2} \left(\distrof{(\RealDecrTrial{}\ 0\ x)}(m) \right) \, dx \cdot F(m \% 2 = 1) \right] \\
    = & \frac{F(\True)}{2} + \sum_{m \in \mathbb{N}} \left[ \int_0^{1/2} \left(\frac{x^{m}}{m!}-\frac{x^{m+1}}{(m+1)!}\right) \, dx \cdot F(m \% 2 = 1) \right] \\
    = & \frac{F(\True)}{2} + \sum_{m \in \mathbb{N}} \left[ \left(\frac{(1/2)^{m+1}}{(m+1)!}-\frac{(1/2)^{m+2}}{(m+2)!}\right) \cdot F(m \% 2 = 1) \right] \\
    = & \frac{F(\True)}{2} - \frac{F(\True)}{2} + \sum_{m \in \mathbb{N}} \left[ \left(\frac{(1/2)^{m}}{m!}-\frac{(1/2)^{m+1}}{(m+1)!}\right) \cdot F(m \% 2 = 0) \right] \\
    = & \sum_{m \in \mathbb{N}} \left[ \left(\frac{(1/2)^{m}}{m!}-\frac{(1/2)^{m+1}}{(m+1)!}\right) \cdot \iverbr{m \% 2 = 0} \right] \cdot F(\True) \\
      & \quad + \sum_{m \in \mathbb{N}} \left[ \left(\frac{(1/2)^{m}}{m!}-\frac{(1/2)^{m+1}}{(m+1)!}\right) \cdot \iverbr{m \% 2 = 1} \right] \cdot F(\False) \\
    = & \sum_{m \in \mathbb{N}} \left[ \left(\frac{(-1/2)^{m}}{m!}\right) \right] \cdot F(\True) + \left(1 - \sum_{m \in \mathbb{N}} \left[ \left(\frac{(-1/2)^{m}}{m!}\right) \right] \right)\cdot F(\False) \\
    = & e^{-1/2} F(\True) + \left(1 - e^{-1/2} \right) F(\False) \\
\end{align*}
\end{small}

\subsubsection{Gaussian Integer Part}

\begin{itemize}
  \item For the integer sample from $\GeoTrial{}$, use
    \begin{align*}
      g(F, k) \eqdef e^{-k(k-1)/2} \cdot h(F, k, \True) + (1-e^{-k(k-1)/2}) \cdot h(F, k, \False)
    \end{align*}
  \item For the boolean sample from $\IterTrial{}$ use
    \begin{align*}
      h(F, k, b) \eqdef \iverbr{b} \cdot F(k) + \iverbr{\neg b} \cdot \sum_{j \in \mathbb{N}} \frac{e^{-j^2/2}}{\normof{\GaussInt}} \cdot F(j)
    \end{align*}
\end{itemize}

\paragraph{Expectation Preservation}

We seek to show that $\sum_{k \in \mathbb{N}} \frac{e^{-k^2/2}}{\normof{\GaussInt}} \cdot F(k) = \sum_{k \in \mathbb{N}} \left(e^{-1/2}\right)^k \cdot (1 - e^{-1/2}) \cdot g(F, k)$. First, we calculate
\begin{small}
\begin{align*}
    & \sum_{k \in \mathbb{N}} \left(e^{-1/2}\right)^k \cdot (1 - e^{-1/2}) \cdot g(F, k) \\
  = & \sum_{k \in \mathbb{N}} \left(e^{-1/2}\right)^k \cdot (1 - e^{-1/2}) \cdot \left( e^{-k(k-1)/2} \cdot F(k) + (1-e^{-k(k-1)/2}) \cdot \sum_{j \in \mathbb{N}} \left[\frac{e^{-j^2/2}}{\normof{\GaussInt}} \cdot F(j)\right]\right) \\
  = & \sum_{k \in \mathbb{N}} \left(e^{-k/2}  - e^{-(k+1)/2}\right) \cdot \left( e^{-k(k-1)/2} \cdot F(k) + (1-e^{-k(k-1)/2}) \cdot \sum_{j \in \mathbb{N}} \left[\frac{e^{-j^2/2}}{\normof{\GaussInt}} \cdot F(j)\right]\right) \\
  = & \sum_{k \in \mathbb{N}} \left(e^{-k/2} \cdot e^{-k(k-1)/2} \cdot F(k) \right) - \sum_{k \in \mathbb{N}} \left(e^{-(k+1)/2} \cdot e^{-k(k-1)/2} \cdot F(k) \right) \\
    & \quad + \sum_{k \in \mathbb{N}} \left(e^{-k/2} \cdot (1-e^{-k(k-1)/2}) \cdot \sum_{j \in \mathbb{N}} \left[\frac{e^{-j^2/2}}{\normof{\GaussInt}} \cdot F(j)\right]\right) \\
    & \quad - \sum_{k \in \mathbb{N}} \left(e^{-(k+1)/2} \cdot (1-e^{-k(k-1)/2})\cdot \sum_{j \in \mathbb{N}} \left[\frac{e^{-j^2/2}}{\normof{\GaussInt}} \cdot F(j)\right]\right) \\
  = & \sum_{k \in \mathbb{N}} \left(e^{-k^2/2} \cdot F(k) \right) - \sum_{n \in \mathbb{N}} \left(e^{-(k^2+1)/2} \cdot F(k) \right) \\
    & \quad + \sum_{k \in \mathbb{N}} \left[\frac{e^{-k^2/2}}{\normof{\GaussInt}} \cdot F(k)\right] \cdot \sum_{j \in \mathbb{N}} \left(e^{-j/2} \cdot (1-e^{-j(j-1)/2})\right) \\
    & \quad - \sum_{k \in \mathbb{N}} \left[\frac{e^{-k^2/2}}{\normof{\GaussInt}} \cdot F(k)\right] \cdot \sum_{j \in \mathbb{N}} \left(e^{-(j+1)/2} \cdot (1-e^{-j(j-1)/2}) \cdot \right) \\
\end{align*}
\end{small}
Combining sums, we can then show the original equality termwise:
\begin{small}
\begin{align*}
      & e^{-k^2/2} - e^{-(k^2+1)/2} + \frac{e^{-k^2/2}}{\normof{\GaussInt}} \sum_{j \in \mathbb{N}} \left(e^{-j/2} \cdot (1-e^{-j(j-1)/2})\right) \\
    & \quad - \frac{e^{-k^2/2}}{\normof{\GaussInt}} \cdot \sum_{j \in \mathbb{N}} \left(e^{-(j+1)/2} \cdot (1-e^{-j(j-1)/2}) \right) \\
    = & e^{-k^2/2} - e^{-(k^2+1)/2} + (1-e^{-1/2}) \cdot \frac{e^{-k^2/2}}{\normof{\GaussInt}} \sum_{j \in \mathbb{N}} \left(e^{-j/2} \cdot (1-e^{-j(j-1)/2})\right) \\
    = & e^{-k^2/2} - e^{-(k^2+1)/2} + (1-e^{-1/2}) \cdot \frac{e^{-k^2/2}}{\normof{\GaussInt}} \left[  \sum_{j \in \mathbb{N}} e^{-j/2} - \sum_{j \in \mathbb{N}} e^{-j^2/2} \right] \\
    = & e^{-k^2/2} - e^{-(k^2+1)/2} + (1-e^{-1/2}) \cdot \frac{e^{-k^2/2}}{\normof{\GaussInt}} \left[ \frac{1}{1-e^{-1/2}} - \normof{\GaussInt} \right] \\
    = & \frac{e^{-k^2/2}}{\normof{\GaussInt}} + e^{-k^2/2} - e^{-(k^2+1)/2} + e^{-(k^2+1)/2} - e^{-k^2/2} \\
    = & \frac{e^{-k^2/2}}{\normof{\GaussInt}}
\end{align*}
\end{small}

\subsubsection{Gaussian Helper}

The credit arithmetic for $\IntChoice$ and $\FractionReject$ are trivial.
The credit arithmetic for $\Selector$ is almost identical to the analysis for $\SelectorZ$, which we will outline here.
The closed form for $\SelectorZ$ that we seek to establish is
\begin{equation*}
  \distrof{(\SelectorZ\ k\ x\ y\ N)}(n) \eqdef \iverbr{N \le n} \cdot \left(\frac{y^{n-N}}{(n-N)!} \left(\frac{2k+x}{2k+2}\right)^{n-N} - \frac{y^{n-N+1}}{(n-N+1)!} \left(\frac{2k+x}{2k+2}\right)^{n-N+1} \right)
\end{equation*}

\begin{itemize}
  \item For the uniform deviate sample from $\eunif{}$, use
    \begin{align*}
     &  g(F, x, k, y, N, z) \eqdef \\
     & \quad \quad \iverbr{y < z} \cdot F(N) \\
     & \quad \quad + \iverbr{z \le y} \cdot \left[\frac{2-x}{2k+2} \cdot h(F, x, k, y, N, z, \True) + \frac{2k+x}{2k+2} h(F, x, k, y, N, z, \False) \right]
    \end{align*}
   \item For the boolean sample from $\FractionReject{}$, use
    \begin{align*}
      h(F, x, k, y, N, z, b) \eqdef \iverbr{b} \cdot F(N) + \iverbr{\neg b} \cdot \sum_{j \in \mathbb{N}} \distrof{(\SelectorZ\ k\ x\ z\ N+1)}(j) \cdot F(j)
    \end{align*}
\end{itemize}

\paragraph{Expectation Preservation}
We seek to show that $\sum_{n \in \mathbb{N}} \distrof{(\SelectorZ\ k\ x\ y\ N)}(n) \cdot F(n) = \int_0^1 g(F, x, k, y, N, z) \, dz$. We calculate with the right-hand side of this equation.
\begin{small}
\begin{align*}
    & \int_0^1 g(F, x, k, y, N, z) \, dz \\
  = & \int_0^1 \iverbr{y < z} \cdot F(N) + \iverbr{z \le y} \cdot \left[\frac{2-x}{2k+2} \cdot F(N) + \frac{2k+x}{2k+2} \left(\sum_{j \in \mathbb{N}} \distrof{(\SelectorZ\ k\ x\ z\ N+1)}(j) \cdot F(j) \right) \right] \, dz \\
  = & \left(1 - y + y \cdot \frac{2-x}{2k+2} \right) \cdot F(N) + \int_0^1  \left[ \iverbr{z \le y} \cdot \frac{2k+x}{2k+2} \left(\sum_{j \in \mathbb{N}} \distrof{(\SelectorZ\ k\ x\ z\ N+1)}(j) \cdot F(j) \right) \right] \, dz \\
  = & \left(1 - y + y \cdot \frac{2-x}{2k+2} \right) \cdot F(N)
     + \frac{2k+x}{2k+2} \sum_{j \in \mathbb{N}} \int_0^y \left(\distrof{(\SelectorZ\ k\ x\ z\ N+1)}(j) \right) \, dz \cdot F(j) \\
\end{align*}
\end{small}

Now, compare coefficients on $F$.
For $n < N$, the Iverson brackets ensure every coefficient on $F(n)$ is zero.
Comparing coefficients on $F(N)$, the Iverson bracket in $\left(\distrof{(\SelectorZ\ k\ x\ z\ N+1)}(j) \right)$ eliminates the rightmost term from the right-hand side, leaving
\begin{small}
\begin{align*}
    \distrof{(\SelectorZ\ k\ x\ y\ N)}(N)
  & = 1 - y + y \cdot \frac{2-x}{2k+2} \\
& = 1 - y \left (1 - \frac{2-x}{2k+2}\right) \\
  & = 1 - y \left (\frac{2k+x}{2k+2}\right) \\
\end{align*}
\end{small}

Lastly, we compare coefficients for $n > N$.
\begin{small}
\begin{align*}
 & \distrof{(\SelectorZ\ k\ x\ y\ N)}(n) \\
 = & \frac{2k+x}{2k+2} \int_0^y \left(\distrof{(\SelectorZ\ k\ x\ z\ N+1)}(n) \right) \, dz \\
= & \frac{2k+x}{2k+2} \int_0^y \left(\frac{y^{n-N-1}}{(n-N-1)!} \left(\frac{2k+x}{2k+2}\right)^{n-N-1} - \frac{z^{n-N}}{(n-N)!} \left(\frac{2k+x}{2k+2}\right)^{n-N} \right) \, dz \\
 = & \int_0^y \left(\frac{y^{n-N-1}}{(n-N-1)!} \left(\frac{2k+x}{2k+2}\right)^{n-N} - \frac{z^{n-N}}{(n-N)!} \left(\frac{2k+x}{2k+2}\right)^{n-N+1} \right) \, dz \\
 = & \frac{y^{n-N}}{(n-N)!} \left(\frac{2k+x}{2k+2}\right)^{n-N} - \frac{y^{n-N+1}}{(n-N+1)!} \left(\frac{2k+x}{2k+2}\right)^{n-N+1}
\end{align*}
\end{small}

\subsubsection{One-sided Gaussian}

Recall that this is a distribution over $\mathbb{N} \times [0,1]$, so $F : \mathbb{N} \to [0,1] \to \mathbb{R}^{\ge 0}$.

\begin{itemize}
  \item For the integer sample from $\GaussInt{}$, use
    \begin{align*}
      f(F, k) \eqdef \int_0^1 g(F, k, x) \, dx
    \end{align*}
  \item For the uniform deviate sample from $\eunif{}$, use
    \begin{align*}
      g(F, k, x) \eqdef e^{-x(2k+x)/2} h(F, x, y, \True) + (1-e^{-x(2k+x)/2}) h(F, x, y, \False)
    \end{align*}
   \item For the boolean sample from $\IterTrial{}$, use
    \begin{align*}
      h(F, x, k, b) \eqdef \iverbr{b} \cdot F(k, x) + \iverbr{\neg b} \cdot \sum_{j \in \mathbb{N}} \int_0^1 \frac{e^{-(j+w)^2/2}}{\normof{\Gauss}} \cdot F(j, w) \, dw
    \end{align*}
\end{itemize}

\paragraph{Expectation Preservation}
\begin{small}
\begin{align*}
    & \sum_{k \in \mathbb{N}} \int_0^1 \frac{e^{-(k+x)^2/2}}{\normof{\Gauss}} \cdot F(k, x) \, dx \\
  = & \sum_{k \in \mathbb{N}} \frac{e^{-k^2/2}}{\normof{\GaussInt}} \int_0^1 e^{-x(2k+x)/2} F(k, x) + (1-e^{-x(2k+x)/2}) \left[\sum_{j \in \mathbb{N}} \int_0^1 \frac{e^{-(j+w)^2/2}}{\normof{\Gauss}} \cdot F(j, w) \, dw \right] \, dx \\
  = & \sum_{k \in \mathbb{N}} \frac{e^{-k^2/2}}{\normof{\GaussInt}} \int_0^1 e^{-x(2k+x)/2} F(k, x) \, dx \\
  = & + \sum_{k \in \mathbb{N}} \frac{e^{-k^2/2}}{\normof{\GaussInt}} \int_0^1 (1-e^{-x(2k+x)/2}) \left[\sum_{j \in \mathbb{N}} \int_0^1 \frac{e^{-(j+w)^2/2}}{\normof{\Gauss}} \cdot F(j, w) \, dw \right] \, dx \\
  = & \sum_{k \in \mathbb{N}} \int_0^1 \frac{e^{-k^2/2}}{\normof{\GaussInt}} e^{-x(2k+x)/2} F(k, x) \, dx + \\
  = & \sum_{k \in \mathbb{N}} \int_0^1 \sum_{j \in \mathbb{N}} \frac{e^{-j^2/2}}{\normof{\GaussInt}} \left[\int_0^1 (1-e^{-w(2j+w)/2}) \frac{e^{-(k+x)^2/2}}{\normof{\Gauss}} \, dw \right] \cdot F(k, x) \, dx \\
\end{align*}
\end{small}
It suffices to show the integrand and summand are equal, allowing us to cancel $F$.
\begin{small}
\begin{align*}
    & \frac{e^{-(k+x)^2/2}}{\normof{\Gauss}} \\
  = & \frac{e^{-k^2/2}}{\normof{\GaussInt}} e^{-x(2k+x)/2} + \sum_{j \in \mathbb{N}} \frac{e^{-j^2/2}}{\normof{\GaussInt}} \left[\int_0^1 (1-e^{-w(2j+w)/2}) \frac{e^{-(k+x)^2/2}}{\normof{\Gauss}} \, dw \right] \\
  = & \frac{e^{-(k+x)^2/2}}{\normof{\GaussInt}} + \frac{e^{-(k+x)^2/2}}{\normof{\Gauss} \cdot \normof{\GaussInt}} \cdot \sum_{j \in \mathbb{N}} e^{-j^2/2} \left[\int_0^1 (1-e^{-w(2j+w)/2}) \, dw \right] \\
  = & \frac{e^{-(k+x)^2/2}}{\normof{\GaussInt}} + \frac{e^{-(k+x)^2/2}}{\normof{\Gauss} \cdot \normof{\GaussInt}} \cdot \left(\sum_{j \in \mathbb{N}} e^{-j^2/2} - e^{-j^2/2} \sum_{j \in \mathbb{N}} \int_0^1 e^{-w(2j+w)/2} \, dw \right)\\
  = & \frac{e^{-(k+x)^2/2}}{\normof{\GaussInt}} + \frac{e^{-(k+x)^2/2}}{\normof{\Gauss} \cdot \normof{\GaussInt}} \cdot \left(\sum_{j \in \mathbb{N}} e^{-j^2/2} - \sum_{j \in \mathbb{N}} \int_0^1 e^{-(w+j)^2/2} \, dw \right)\\
  = & \frac{e^{-(k+x)^2/2}}{\normof{\GaussInt}} + \frac{e^{-(k+x)^2/2}}{\normof{\Gauss} \cdot \normof{\GaussInt}} \cdot \left(\normof{\GaussInt} - \normof{\Gauss} \right)\\
  = & \frac{e^{-(k+x)^2/2}}{\normof{\Gauss}} \left( \frac{\normof{\Gauss}}{\normof{\GaussInt}} + \frac{\normof{\GaussInt} - \normof{\Gauss}}{\normof{\GaussInt}} \right) \\
  = & \frac{e^{-(k+x)^2/2}}{\normof{\Gauss}} \\
\end{align*}
\end{small}

\subsubsection{Negative Exponential}

The negative exponential is also a distribution over $\mathbb{N} \times [0,1]$, so $F : \mathbb{N} \to [0,1] \to \mathbb{R}^{\ge 0}$.
We seek to show that $\NegExp \ N$ is distributed as
\begin{equation*}
  \distrof{(\NegExp\ N)}(k, x) \eqdef \iverbr{N \le k} e^{-(x+k-N)}
\end{equation*}

\begin{itemize}
  \item For the uniform deviate sample from $\eunif{}$ use
    \begin{align*}
      g(F, N, x) \eqdef \sum_{n \in \mathbb{N}} \left(\frac{x^n}{n!} - \frac{x^{n+1}}{(n+1)!}\right) \cdot h(F, N, x, n)
    \end{align*}
  \item For the integer sample from $\RealDecrTrial{}$, use
    \begin{align*}
      & h(F, N, x, k) \eqdef \\
      & \quad \iverbr{k \% 2 = 0} \cdot F(N, x) + \iverbr{k \% 2 = 1} \cdot \sum_{j \in \mathbb{N}}\int_0^1 \iverbr{N+1 \le j} \cdot e^{-(j-N-1+y)} \cdot F(j, y) \, dy
    \end{align*}
\end{itemize}

\paragraph{Expectation Preservation}
\begin{small}
\begin{align*}
    & \sum_{k \in \mathbb{N}}\int_0^1 \iverbr{N \le k} e^{-(x+k-N)} \cdot F(k, x) \\
  = & \int_0^1 g(F, N, x) \, dx \\
  = & \int_0^1 \sum_{k \in \mathbb{N}} \left(\frac{x^k}{k!} - \frac{x^{k+1}}{(k+1)!}\right) \\
    & \quad \cdot \left( \iverbr{k \% 2 = 0} \cdot F(N, x) + \iverbr{k \% 2 = 1} \cdot \sum_{j \in \mathbb{N}}\int_0^1 \iverbr{N+1 \le j} \cdot e^{-(j-N-1+y)} \cdot F(j, y) \, dy \right) \, dx \\
  = & \int_0^1 \sum_{k \in \mathbb{N}} \left(\frac{x^k}{k!} - \frac{x^{k+1}}{(k+1)!}\right) \cdot \iverbr{k \% 2 = 0} \cdot F(N, x) \, dx \\
    & \quad + \int_0^1 \sum_{k \in \mathbb{N}} \left(\frac{x^n}{n!} - \frac{x^{n+1}}{(n+1)!}\right) \cdot \iverbr{k \% 2 = 1} \cdot \sum_{j \in \mathbb{N}}\int_0^1 \iverbr{N+1 \le j} \cdot e^{-(j-N-1+y)} \cdot F(j, y) \, dy \, dx \\
  = & \int_0^1 \sum_{k \in \mathbb{N}} \left(\left(\frac{x^k}{k!} - \frac{x^{k+1}}{(k+1)!}\right) \cdot \iverbr{k \% 2 = 0} \right) \cdot F(N, x) \, dx \\
    & \quad + \left( \int_0^1 \sum_{k \in \mathbb{N}} \left(\frac{x^n}{n!} - \frac{x^{n+1}}{(n+1)!}\right) \cdot \iverbr{k \% 2 = 1} \, dx \right) \cdot \\
    & \quad\quad \left(\sum_{j \in \mathbb{N}}\int_0^1 \iverbr{N+1 \le j} \cdot e^{-(j-N-1+y)} \cdot F(j, y) \, dy \right) \\
  = & \int_0^1 e^{-x} \cdot F(N, x) \, dx \\
  & \quad + \left( \int_0^1 (1-e^{-x}) \, dx \right) \cdot \left(\sum_{j \in \mathbb{N}}\int_0^1 \iverbr{N+1 \le j} \cdot e^{-(j-N-1+y)} \cdot F(j, y) \, dy \right) \\
  = & \int_0^1 e^{-x} \cdot F(N, x) \, dx + \frac{1}{e} \cdot \left(\sum_{k \in \mathbb{N}}\int_0^1 \iverbr{N+1 \le k} \cdot e^{-(k-N-1+x)} \cdot F(k, x) \, dx \right) \\
  = & \int_0^1 e^{-x} \cdot F(N, x) \, dx + \left(\sum_{k \in \mathbb{N}}\int_0^1 \iverbr{N+1 \le k} \cdot e^{-(k-N+x)} \cdot F(k, x) \, dx \right) \\
  = & \sum_{k \in \mathbb{N}}\int_0^1 \iverbr{N \le k} \cdot e^{-(k-N+x)} \cdot F(k, x) \, dx \\
\end{align*}
\end{small}
 }{}

\appendix

\end{document}